\documentclass[structabstract]{aa}  

\usepackage{graphicx}
\usepackage{textcomp}
\usepackage{longtable,lscape}
\usepackage[varg]{txfonts}
\usepackage{mathrsfs}
\usepackage{natbib}
\usepackage[colorlinks, citecolor={blue}]{hyperref}

\begin{document}

   \title{Effect of local and large-scale environments on nuclear activity and star formation}

   \author{M.~Argudo-Fern\'andez\inst{1,2}
          \and
          S.~Shen\inst{1,3}
          \and 
          J.~Sabater\inst{4}
          \and
          S. Duarte Puertas\inst{5}
          \and
          S. Verley\inst{6,7}
          \and
          X. Yang\inst{8,9}}

   \institute{Key Laboratory for Research in Galaxies and Cosmology, Shanghai Astronomical Observatory, Chinese Academy of Sciences, 80 Nandan Road, Shanghai, China, 200030
         \and
             Universidad de Antofagasta, Unidad de Astronom\'ia, Facultad Cs. B\'asicas, Av. U. de Antofagasta 02800, Antofagasta, Chile
         \and    
             Key Lab for Astrophysics, Shanghai, 200234, China
         \and   
             Institute for Astronomy, University of Edinburgh, Edinburgh EH9 3HJ, UK
         \and         
             Instituto de Astrof\'isica de Andaluc\'ia (CSIC) Apdo. 3004, 18080 Granada, Spain
         \and
             Departamento de F\'isica Te\'orica y del Cosmos, Universidad de Granada, 18071 Granada, Spain
         \and
             Instituto Universitario Carlos I de F\'isica Te\'orica y Computacional, Universidad de Granada, 18071 Granada, Spain
         \and
             Center for Astronomy and Astrophysics, Shanghai Jiao Tong University, Shanghai 200240, China
         \and 
             IFSA Collaborative Innovation Center, Shanghai Jiao Tong University, Shanghai 200240, China}

   \date{Received 2 February 2016; accepted 18 May 2016}

 
\abstract
{Active galactic nuclei (AGN) is one of the main drivers for transition from star-forming disk to passive spheroidal galaxies. However, the role of large-scale environment versus one-on-one interactions in triggering different types of AGN is still uncertain. We present a statistical study of the prevalence of the nuclear activity in isolated galaxies and physically bound isolated pairs.}
{For the purpose of this study we considered optically and radio selected nuclear activity types. We aim to assess the effect of one-on-one interaction on the fraction of AGN and the role of their large-scale environment.}
{To study the effect of one-on-one interaction on the fraction of AGN in isolated galaxy pairs, we compare with a sample of isolated galaxies homogeneously selected under the same isolation criterion. We examine the effect of the large-scale environment by comparing with control samples of single galaxies and galaxy pairs. To quantify the effects of local and large-scale environments we use the tidal strength parameter.}
{In general we found no difference in the prevalence of optical AGN for the considered samples. For massive galaxies, the fraction of optical AGN in isolated galaxies is slightly higher than that in control samples. Also the fraction of passives in high mass isolated galaxies is smaller than in any other sample. Generally, there is no dependence on optical nuclear activity with local environment. On the other hand, we found evidence that radio AGN are strongly affected by the local environment.}
{Optical AGN phenomenon is related to cold gas accretion, while radio AGN is related to hot gas accretion. In this context, there is more cold gas, fuelling the central optical AGN, in isolated systems. Our results are in agreement with a scenario where cold gas accretion by secular evolution is the main driver of optical AGN, while hot gas accretion and one-on-one interactions are the main drivers of radio AGN activity.}


   \maketitle

%


\section{Introduction}

The environment in which a galaxy resides plays an important role on its formation and evolution. Galaxies suffer intrinsic and secular evolution processes (i.e. nature processes), but they are also exposed to the influences of their local and large-scale environments (i.e. nurture processes) \citep{2015MNRAS.451..888C}. Starting from the well known morphology-density relation \citep{1980ApJ...236..351D}, other properties such as stellar mass, size, or colour are influenced by environmental processes \citep{2010ApJ...721..193P,2011MNRAS.411..929G,2012MNRAS.419L..14C,2012ApJ...757....4P,2014A&ARv..22...74B}. Moreover, nuclear activity is somehow affected by galaxy environment \citep{2004MNRAS.353..713K,2009ApJ...699.1679C,2013MNRAS.430..638S}.

Active galactic nuclei (AGN) also plays an important role in galaxy formation and evolution. Observations have led to the interpretation that the main driver for transition from star-forming disk to passive spheroidal galaxies is through AGN \citep{2003MNRAS.346.1055K}. A galaxy is revealed as an AGN when its central black hole (BH) grows through mass accretion, liberating in the process a huge amount of energy \citep[see][]{1964ApJ...140..796S,1969Natur.223..690L,1982MNRAS.200..115S,1999MNRAS.308L..39F,2012NewAR..56...93A}. There is some evidence that galaxies are more likely to host an AGN when they are interacting with a neighbour \citep[e.g.][]{2011MNRAS.418.2043E}. Since the most evolved and massive galaxies reside in clusters at the present day \citep{1958ApJS....3..211A,2004ApJ...600..681B}, the environment associated to the large-scale structure (LSS) is therefore also likely to have a significant effect on the triggering of AGN activity. However the physical processes responsible for triggering AGN, as well as the role of the environment on the fuelling of BHs, are still uncertain.  

Is AGN activity connected to environment? There has been some disagreement in the literature regarding the connection between environment and nuclear activity. Differences are mainly due to sample selections and the diverse definitions of environment. The statistical studies developed by \citet{2013MNRAS.430..638S,2015MNRAS.447..110S} suggest that large-scale environment and galaxy interactions play a fundamental but indirect role in AGN activity (by influencing the gas supply) and that the dependence on AGN luminosity is minimal. On the other hand, the results from deep images by \citet{2015ApJ...804...34H} suggest that luminous AGN activity is associated with galaxy merging. These differences could be explained since the results of the first studies, based on Sloan Digital Sky Survey \citep[SDSS,][]{2000AJ....120.1579Y,2002AJ....124.1810S} seventh data release \citep[DR7;][]{2009ApJS..182..543A} data, might consider merger systems as single galaxies and also be affected by the well know fibre-collision problem for close objects \citep{2002AJ....124.1810S}. 

Even if it can be difficult reconciling the results from one study with the results from another study, there is some consensus that secular processes may be much more important in driving the black hole growth than previously assumed \citep{2015aska.confE..83M,2015MNRAS.447..110S}, where the request for AGN triggering is an abundant supply of central cold gas, regardless of its origin. In the case of high luminosity AGN, major mergers appear to be the main driver \citep{2011MNRAS.418.2043E,2012MNRAS.419..687R,2015MNRAS.452..774K,2014ApJ...788..140M,2014MNRAS.441.1297S,2015ApJ...806..147C}. 

There are two modes of AGN activity: the 'quasar-mode' \citep{1973A&A....24..337S}, or 'cold-mode', and the 'radio-mode' \citep{1979MNRAS.188..111H}, or 'hot-mode'. In 'cold-mode' the AGN radiates across a broad multi-wavelength range (including optical and radio), while in 'hot-mode' AGN are mainly detectable in radio, due to the emission of their jet. The two accretion modes have different associated feedback effects in the host galaxy \citep{2003MNRAS.346.1055K,2005MNRAS.362...25B}, and they are powered by accretion of different material that can be connected to the environment \citep{2007MNRAS.376.1849H,2008A&A...490..893T}, but the precise origin of these differences effects remains unclear \citep{2012MNRAS.421.1569B}. 

The aim of the present study is to accurately measure the fraction of optical and radio AGN activity that is triggered by external or internal processes. With the purpose of determining the importance of secular evolution versus one-on-one interactions in triggering different types of AGN activity, we select samples of both isolated galaxies and isolated pairs. In this study we focus on isolated systems because any difference in the AGN fraction would be directly related to the addition of one companion. Also, by carefully selecting control samples of galaxy pairs and single galaxies, we go one step further and explore the effect of the large-scale environment versus one-on-one interactions on the AGN prevalence. 

This study is organised as follows. In Sect.~\ref{Sec:data} we describe the samples used in this work as well as the selected AGN classification methods and the parameters used to quantify the environment. We present our results in Sect.~\ref{Sec:res} and the associated discussion in Sect.~\ref{Sec:dis}. Finally, a summary and the main conclusions of the study are presented in Sect.~\ref{Sec:con}. Throughout the study, a cosmology with $\Omega_{\Lambda 0} = 0.7$, $\Omega_{\rm{m} 0} = 0.3$, and $H_{0}=70$\,km\,s$^{-1}$\,Mpc$^{-1}$ is assumed.

\section{Data and methodology} \label{Sec:data}

To study the effect of one-on-one interaction on the fraction of AGN, we compare isolated galaxy pairs with a sample of isolated galaxies homogeneously selected under the same isolation criterion. To understand the effect of the large-scale environment, we select control samples composed of single galaxies and galaxy pairs that can be found in any environment, from clusters or groups to voids. 

\subsection{Isolated galaxies and isolated pairs}

We use the sample of isolated galaxies and isolated pairs compiled by \citet{2015A&A...578A.110A} from the SDSS-DR10 \citep{2014ApJS..211...17A}. Isolated galaxies and central galaxies in the pairs (the brightest of the pair by definition) are selected in a volume limited sample redshift range $0.005 \leq z \leq 0.080$, with $11 \leq m_{r} \leq 15.7$, where $m_{r}$ is the SDSS model magnitude in the $r$-band. This criterion allows the second galaxy in the pair to be at least 2 orders of magnitude fainter within the range of spectroscopic completeness of the SDSS main galaxy sample at $m_{r,\rm{Petrosian}}~<~17.77$\,mag \citep{2002AJ....124.1810S}.

\cite{2015A&A...578A.110A} used a three dimensional isolation criterion, based on projected distances on the sky and redshift. The systems are isolated with no neighbours in a volume of 1\,Mpc projected distance within a line-of-sight velocity difference of $\Delta\,\varv~\leq~500$\,km\,s$^{-1}$. After defining isolation, they followed a similar method as in \citet{2014A&A...564A..94A} to identify the physically bound isolated pairs. For the central galaxy in isolated pairs, \citet{2015A&A...578A.110A} found an over-density of neighbours in the 2D distribution of $\Delta\,\varv$ and distance $d$. This over-density indicates that those neighbours are likely to be physically connected. The distribution of $\Delta\,\varv$ for those neighbours follows a Gaussian distribution. The neighbour galaxies within $\Delta\,\varv \leq 2 \sigma$\,(160\,km\,s$^{-1}$) show also a tendency to be located within the first 450\,kpc from the central galaxy. Conversely, neighbour galaxies at higher $\Delta\,\varv$ and $d$ would be associated to the underlying large-scale distribution of galaxies \citep{2015A&A...578A.110A}. They found 3702 isolated galaxies, hereafter SIG (SDSS-based Isolated Galaxies), and 1240 isolated pairs physically bound at projected distances up to $d~\leq~450$\,kpc within $\Delta\,\varv~\leq~160$\,km\,s$^{-1}$, hereafter SIP (SDSS-based Isolated Pairs). The SIG and SIP samples represent about 11\% and 7\% of the galaxies in the local Universe \citep[$z\leq0.080$;][]{2015A&A...578A.110A}. The average projected distance and $\Delta\,\varv$ of the SIP sample is $d~\simeq~215$\,kpc and $\Delta\,\varv~\simeq~65$\,km\,s$^{-1}$. The average stellar mass\footnote{Stellar masses in the CIG and SIP samples come from the fitting to the spectral energy distribution on the five SDSS bands using the routine kcorrect \citep{2007AJ....133..734B} and the relation between the stellar mass-to-light ratio and color of \citet{2003ApJS..149..289B}.} ratio in isolated pairs ($\rm \frac{M_{\star B}}{M_{\star A}}$, where A corresponds to the central galaxy and B to the faintest galaxy in the pair) is $\rm \frac{M_{\star B}}{M_{\star A}}~\simeq~0.30$, in the range $\rm 0.01~\lesssim~\frac{M_{\star B}}{M_{\star A}}~\lesssim~1.00$ \citep[see][for further details]{2015A&A...578A.110A}.

\begin{figure}[!htbp]
\centering
\includegraphics[width=\columnwidth]{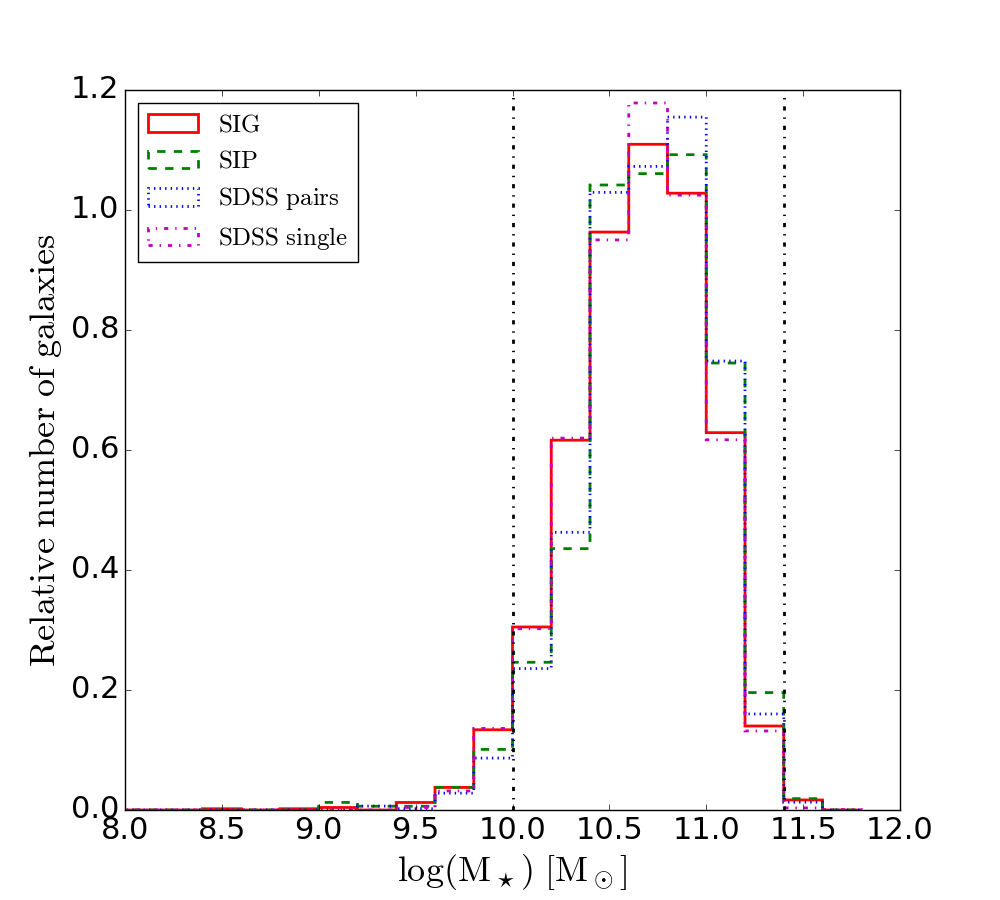}\\
\includegraphics[width=\columnwidth]{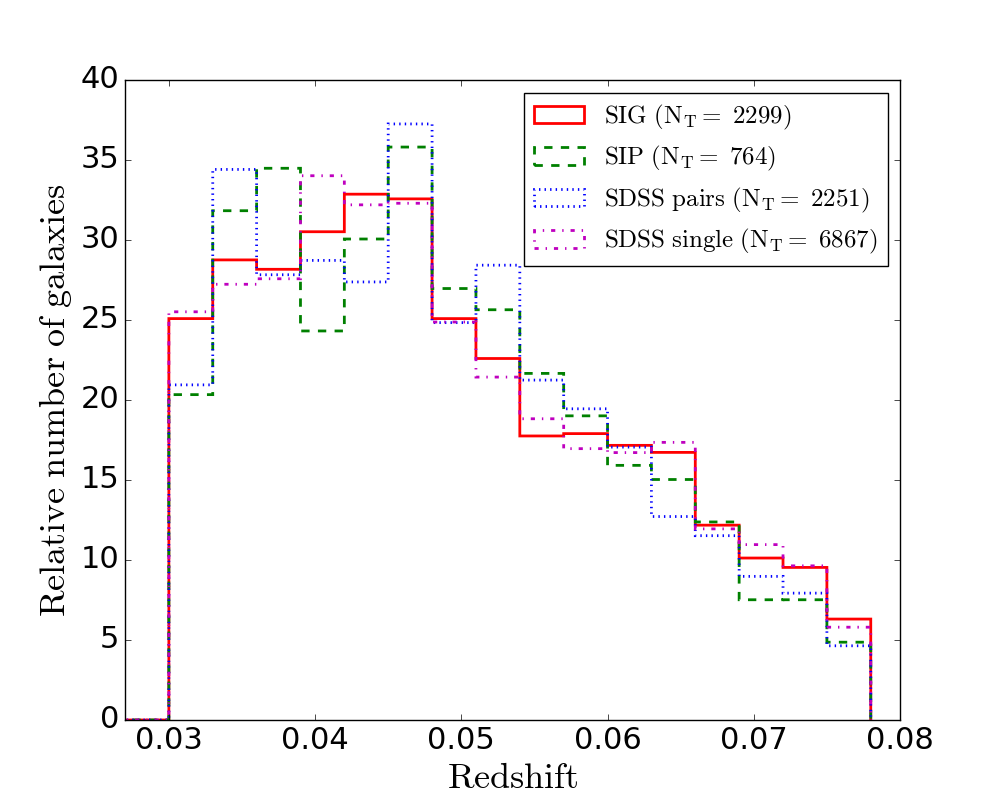}
\caption{{\it Upper panel:} Distribution of the stellar masses of the SIG and SDSS single galaxies (red solid and magenta dash-dotted histograms, respectively), and for central SIP and SDSS pairs (green dashed and blue dotted histograms, respectively). The vertical black dashed lines correspond to the selected stellar mass range for the study at $10.0~\leq~\rm{log}(M_\star)~\leq~11.4~[M_\odot]$. {\it Lower panel:} Distribution of the redshift of the SIG and SDSS single galaxies (red solid and magenta dash-dotted histograms, respectively), and for central SIP and SDSS pairs (green dashed and blue dotted histograms, respectively), for galaxies within the stellar mass limited sample ($N_T$).}
\label{Fig:hist_mass}
\end{figure} 

\subsection{Control galaxy samples}

The control samples are based on the catalogue of groups compiled by \citet{2007ApJ...671..153Y}, which is based on the NYU-VAGC \citep{2005AJ....129.2562B} updated with SDSS-DR7 data. \citet{2007ApJ...671..153Y} developed a halo-based group finder that is optimized for grouping galaxies that reside in the same dark matter halo, including isolated galaxies in small mass haloes. Group members are complete to $M_{r}\,\leq\,-19.5$ absolute magnitude and $z~\leq~0.090$. Nevertheless, some SDSS-DR12 \citep{2015ApJS..219...12A} and LAMOST \citep[Large Sky Area Multi-Object Fiber Spectroscopic Telescope,][]{2012RAA....12..723Z} redshifts are supplied to update the group catalogue \citep{2016RAA....16c...7S}.  

To study the dependence of the large-scale environment on nuclear activity in galaxy pairs, we selected the central (brightest) galaxy in groups composed of two members in \citet{2007ApJ...671..153Y}, hereafter the SDSS pairs sample. Following the same criteria, we selected one-member groups in \citet{2007ApJ...671..153Y}, hereafter the SDSS single sample, as a control sample to compare with SIG galaxies. 

The average projected separation and $\Delta\,\varv$ of SDSS pairs is $d~\simeq~200$\,kpc and $\Delta\,\varv~\simeq~85$\,km\,s$^{-1}$, similar to the SIP sample, where the 90\% of the pairs show projected separations $d~\lesssim~450$\,kpc and $\Delta\,\varv~\lesssim~200$\,km\,s$^{-1}$. The stellar mass\footnote{Stellar masses in the group catalogue were computed fitting the spectral energy distribution on the five SDSS + JHK bands using the routine kcorrect \citep{2007AJ....133..734B} and the relation between the stellar mass-to-light ratio and color of \citet{2003ApJS..149..289B} \citep[see][for further details]{2007ApJ...671..153Y}.} ratio also spans a similar range with an average value $\rm \frac{M_{\star B}}{M_{\star A}}~\simeq~0.23$.

We selected control samples within the same volume limited sample where the isolated systems were defined ($11 \leq m_{r} \leq 15.7$ and $0.005 \leq z \leq 0.080$) to avoid biases. We also removed the galaxies in common with the SIP or SIG catalogues. Following these criteria, the SDSS pairs sample is composed of 3772 pairs (after removing 505 pairs in common with the SIP), and the SDSS single sample is composed of 9,526 galaxies (after removing 2027 galaxies in common with the SIG).

\subsection{Stellar masses and AGN classification} \label{Sec:AGN}

For the purpose of this study we considered optically and radio selected nuclear activity types. In particular, we used published stellar masses and AGN classifications for galaxies in the SDSS-DR7 from \citet{2013MNRAS.430..638S}, hereafter SBA13 classification. 

\defcitealias{2013MNRAS.430..638S}{SBA13}

\citetalias{2013MNRAS.430..638S} used information about total stellar masses from \citet{2003MNRAS.341...33K} and optical AGN classification from BPT diagnostic \citep{1981PASP...93....5B,2003MNRAS.346.1055K}. Data based on optical spectra, i.e. the stellar masses and the corrected emission-line fluxes used to built BPT diagrams, were drawn from the Max Plank Institute for Astrophysics and Johns Hopkins University \citep[MPIA-JHU\footnote{Available at \texttt{http://www.mpa-garching.mpg.de/SDSS/DR7/}};][]{2003MNRAS.341...33K,2004ApJ...613..898T,2007ApJS..173..267S} added value catalogue \citep{2004MNRAS.351.1151B}. 

Radio AGN galaxies in \citetalias{2013MNRAS.430..638S} are considered if they are classified as radio AGN by \citet{2012MNRAS.421.1569B} with a radio luminosity brighter than $L_{1.4~\rm{GHz}}~\approx~10^{23}$\,W\,m$^{-2}$\,Hz$^{-1}$ \citepalias[see][for more details]{2013MNRAS.430..638S}. The radio AGN classification in \citet{2012MNRAS.421.1569B} is based on radio-continuum data from the National Radio Astronomy Observatory (NRAO) Very Large Array (VLA) Sky Survey  \citet[NVSS,][]{1998AJ....115.1693C} and the Faint Images of the Radio Sky at Twenty centimetres \citep[FIRST,][]{1995ApJ...450..559B} data bases, and follows the techniques presented in \citet{2005MNRAS.362....9B}.

Even if we take care on selecting galaxies in the control samples within the same volume limited as isolated galaxies and isolated pairs, it is well known that AGN fraction depends strongly on stellar mass \citep{2003MNRAS.346.1055K,2010ApJ...721..193P,2012ApJ...757....4P}. Henceforth, to have a robust control sample for the comparisons, we selected galaxies in the SDSS single and SDSS pairs samples with similar stellar mass\footnote{Note that stellar masses from \citet{2015A&A...578A.110A} and \citet{2007ApJ...671..153Y} are calculated slightly different, therefore in order to use a consistent source, we consider stellar masses in \citetalias{2013MNRAS.430..638S}.} and redshift distributions than the SIG and SIP samples, respectively (with Kolmogorov-Smirnov p-value greater than 0.80, which ensures that the distributions of the two samples are the same). To have enough numbers of objects in each mass bin for each sample, we also considered galaxies with stellar masses within the range $10.0~\leq~\rm{log}(M_\star)~\leq~11.4~[M_\odot]$ (see the upper panel in Fig.~\ref{Fig:hist_mass}). The final number of galaxies, in the stellar mass range considered in this study, with available \citetalias{2013MNRAS.430..638S} classification is shown in the first row of Table~\ref{tab:samples}.

Due to the number of galaxies in our samples, we only separate nuclear activity into optical AGN, radio AGN, star-forming nuclei (SFN), and passive galaxies (in case that no optical nuclear activity is detected). Optical AGN classification covers transition objects (TO), Seyfert (Seyfert 1 not included), and low-ionization nuclear emission-line region \citep[LINER;][]{1980A&A....87..152H} galaxies. In the case of radio AGN, this includes low-excitation (LERG) and high-excitation (HERG) radio galaxies. The number of galaxies in each sample, classified in each type and subtypes of nuclear activity, is shown in Table~\ref{tab:samples}. Note that \citet{2012MNRAS.421.1569B} classified radio galaxies into HERG and LERG when such classification was possible. Given that there is only one SIP galaxy without classification we rejected this galaxy in the present study. Besides, the three HERG galaxies present in the sample were discarded during the comparisons. Henceforth, with the term radio AGN we will be considering only LERG radio AGN.

Note that, to have a statistically significant number of galaxies, we do not impose any limit on the $[\rm O_{III}]_{5007}$ emission line luminosity. Some low luminosity AGN (usually LINERs) at higher redshifts could be classified as passive if their emission is not strong enough to be detected. However, the overall effect is minimized if the redshift distribution of the samples is relatively similar as it is in our case (see lower panel in Fig.~\ref{Fig:hist_mass}).

Note also that AGN classification based on purely emission-line BPT diagrams are affected by uncertainties \citep{2012A&A...545A..15S}. According to \citet{2016MNRAS.457.2703R}, these uncertainties are specially important for massive main-sequence local galaxies that might be misclassified as passives. We have checked that less than the 4\% of the total number of galaxies in each sample in our study would be affected. Given this number, we do not expect any change in the observed trends for SFN galaxies. The results of \citet{2012A&A...545A..15S} suggest that at least some of these misclassified galaxies could be classified as LINERs or Seyferts if the uncertainties were taken into account. However, given that our samples follow the same classification criteria and a relatively similar distribution in mass and redshift, the possible effect on the comparison between samples will be minimised.

\begin{table}
 \caption{Number of galaxies of each type in each sample.}
 \label{tab:samples}
\centering 
\begin{tabular}{lcccc}
\hline
 Type & SIP & SDSS pairs & SIG & SDSS single  \\
\hline
Total              & 764 & 2251 & 2299 & 6867 \\
\\
Optical AGN        & 387 & 1009 & 1153 & 3027 \\
\,\,\,\,\, LINER   & 152 & 454  & 482  & 1081 \\
\,\,\,\,\, Seyfert & 44  & 96   & 123  & 266  \\
\,\,\,\,\, TO      & 191 & 459  & 548  & 1680 \\
SFN                & 169 & 564  & 587  & 1777 \\
Passive            & 208 & 678  & 559  & 2063 \\
\hline
Radio AGN          & 10  & 39   & 11   & 22   \\
\,\,\,\,\, HERG    & 1   & 2    & 0    & 0    \\
\,\,\,\,\, LERG    & 8   & 37   & 11   & 22   \\
\end{tabular}
\tablefoot{Meaning of the different types: Total -- total number of galaxies in each sample; Optical AGN -- galaxies classified as LINER, Seyfert, or transition objects (TO); SFN -- star forming nuclei galaxies; Passive -- galaxies with no optical nuclear activity; Radio AGN -- galaxies classified as HERG or LERG radio AGN with $L_{1.4~\rm{GHz}}~\geq~10^{23}$\,W\,m$^{-2}$\,Hz$^{-1}$.}
\end{table}

\begin{figure}[!htbp]
\centering
\includegraphics[width=\columnwidth]{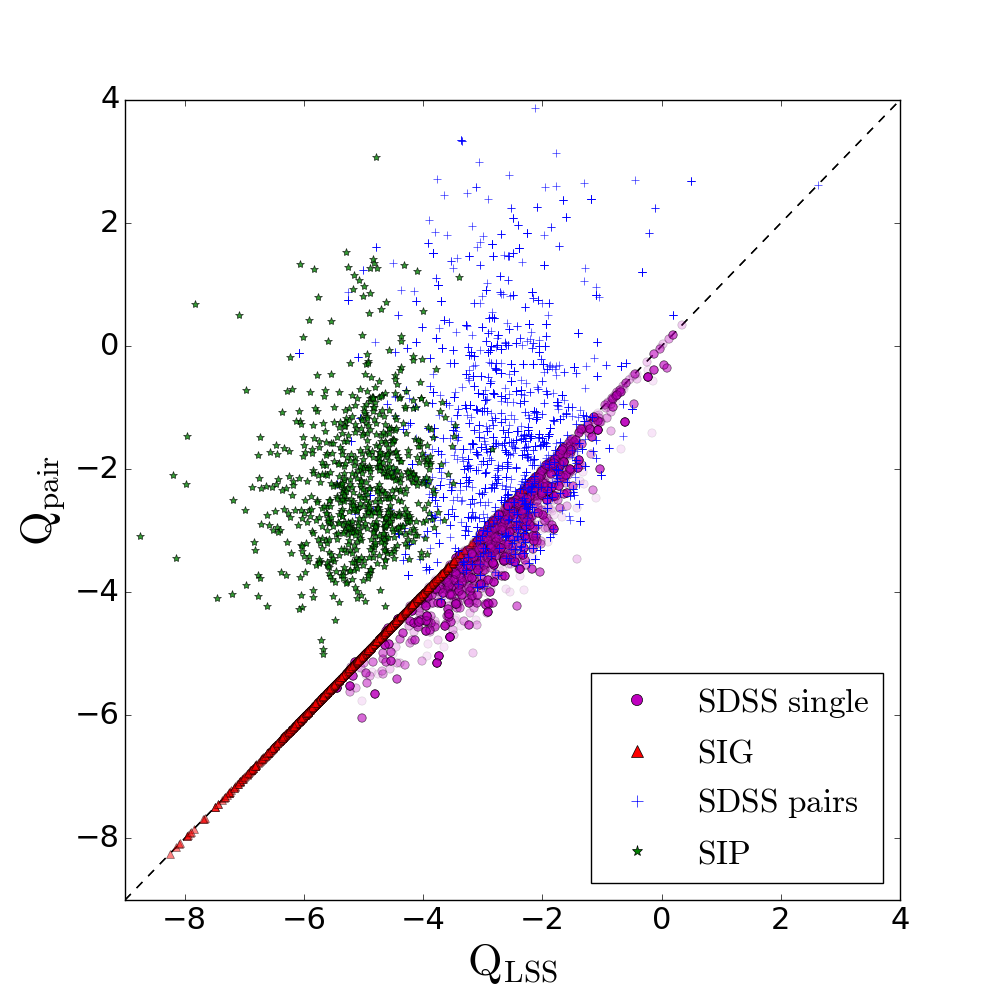}\\
\includegraphics[width=\columnwidth]{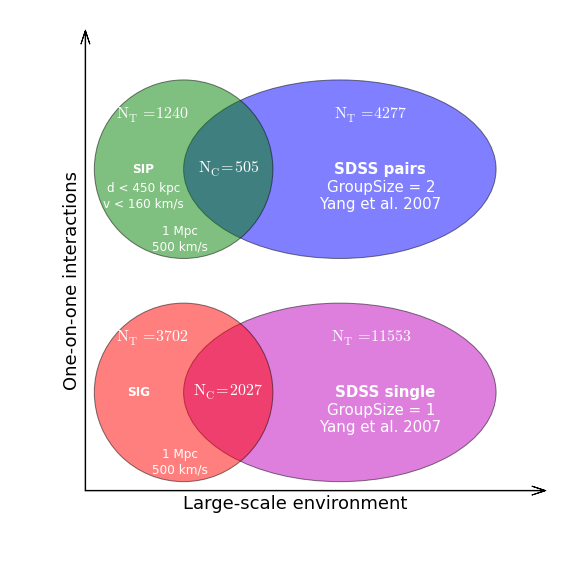}
\caption{{\it Upper panel:} Comparison of $Q_{\rm pair}$ versus $Q_{\rm LSS}$, isolated pairs (green stars), SDSS single galaxies (magenta circles) and SDSS pairs (blue pluses). Note that for isolated galaxies (red triangles) there is not available $Q_{\rm pair}$ but we consider $Q_{\rm pair}~=~Q_{\rm LSS}$ for comparison purposes. The black dashed line represents the line where $Q_{\rm pair}~=~Q_{\rm LSS}$. {\it Lower panel:} Schema of the sample definition, total number of galaxies ($N_T$) in each sample and number of galaxies in common ($N_C$), and environment definition for the SIG and SIP samples (red and green circles, respectively), and for control samples (blue and magenta ellipses for SDSS pairs and SDSS singles respectively). The arrows in the axis indicates the direction to higher values of the tidal strength.}
\label{Fig:Qsamples}
\end{figure}

\subsection{Environmental parameters}

\citet{2015A&A...578A.110A} also provided the isolation degree for isolated galaxies and central galaxies in isolated pairs. They quantified the influence of both, local and large-scale environments, using the tidal strength parameter \citep{2007A&A...472..121V,2013MNRAS.430..638S,2013A&A...560A...9A,2014A&A...564A..94A}.

To study the influence of the large-scale environment on the AGN fraction for isolated galaxies and isolated pairs, we selected the tidal strength exerted by all the galaxies in the LSS up to 5 Mpc, $Q_{\rm LSS}$ (Eq.~\ref{Eq:QLSS}). For isolated pairs, we study the influence of the companion using the local tidal strength $Q_{\rm{pair}}$ (Eq.~\ref{Eq:Qlocal}). 

Galaxies in the LSS are defined within a volume of 5\,Mpc projected distance and line-of-sight velocity difference of $\Delta\,\varv~\leq~500$\,km\,s$^{-1}$. Then, for each galaxy, $i$, in the LSS at a projected distance $d_{LSS_i}$, the total tidal strength on the isolated galaxy (SIG), or the central (brightest) galaxy in isolated pairs (SIP), is:

\begin{equation} \label{Eq:QLSS}
Q_{\rm LSS} \equiv {\rm log} \left(\sum_i {\frac{M_{LSS_i}}{M}} \left(\frac{D}{d_{LSS_i}}\right)^3\right) \quad,
\end{equation} 
where $M$ is the stellar mass and $D$\footnote{$D = 2\alpha r_{90}$, where $r_{90}$, the Petrosian radius containing 90\,\% of the total flux of the galaxy in the $r$-band, is scaled by a factor $\alpha=1.43$ to recover the $D_{25}$ \citep{2013A&A...560A...9A}.} is the estimated diameter of the SIG/SIP galaxy. Stellar masses for galaxies in the LSS ($M_{LSS_i}$) were calculated by fitting the spectral energy distribution, on the five SDSS bands, using the routine kcorrect \citep{2007AJ....133..734B}. 

We quantify the effect of the local environment as the tidal strength affecting the central (brightest) galaxy in a galaxy pair. Then, similarly to the definition of the $Q_{\rm LSS}$ but only considering two galaxies, the local tidal force exerted by the faintest B galaxy on the brightest (central) A galaxy is:  
\begin{equation} \label{Eq:Qlocal}
Q_{\rm pair} \equiv {\rm log} \left(\frac{M_{B}}{M_{A}} \left(\frac{D_{A}}{d_{AB}}\right)^3\right) \quad,
\end{equation} 
where $d_{AB}$ is the projected physical distance between the galaxies of the isolated pair. 

We follow the same methodology to estimate the tidal strength for control samples, i. e. we estimate $Q_{\rm LSS}$ for SDSS single galaxies, and $Q_{\rm pair}$ for SDSS pairs. A scheme of the environment for the four samples is shown in Fig.~\ref{Fig:Qsamples}. For comparison purpose, $Q_{\rm pair}$ for SDSS single galaxies is estimated considering their first nearest neighbour. In the case of the SIG sample, as the nearest neighbour is as least at 1\,Mpc away, the value of the tidal strength exerted by this neighbour is practically the same as the one exerted by the LSS, we therefore consider $Q_{\rm pair}~=~Q_{\rm LSS}$. $Q_{\rm LSS}$ for the control samples is estimated as for SIG and SIP, considering their LSS up to 5\,Mpc. The greater the value of $Q$, the less isolated from external influence the galaxy. Therefore, as it is schematically shown in the lower panel of Fig.~\ref{Fig:Qsamples}, the SIG and SIP samples have the same degree of isolation with respect the LSS, while control samples extend a broader range of large-scale environments. With respect to the local environment, SIG and SDSS singles have a similar range, while the effect is stronger for central galaxies in the SIP and SDSS pairs samples. Then at fixed stellar mass, higher values of $Q_{\rm{pair}}$ are related to closer pairs. Since there is a strong dependence of AGN with stellar mass, it is recommendable to made a separated study in different stellar mass bins \citepalias{2013MNRAS.430..638S}. 

The black dashed line in the upper panel of Fig.~\ref{Fig:Qsamples} corresponds to the line where $Q_{\rm{pair}}$ = $Q_{\rm LSS}$. When a galaxy is located on this line, the contributions by its local and large-scale environment on the total tidal strengths are the same. 
Central galaxies in the SIP sample are located above the line, which means that their tidal strengths are dominated by their close environment. In fact, more than 95\% of the total tidal strength in SIP galaxies is due to the companion galaxy in the pair \citep{2015A&A...578A.110A}. In general SDSS pairs are also located above the line, but there are some pairs below the line. These are pairs located in high density environments, mainly in clusters, surrounded by massive and nearby galaxies which are likely affecting their evolution.

\begin{figure*}[!htbp]
\centering
\includegraphics[width=\textwidth]{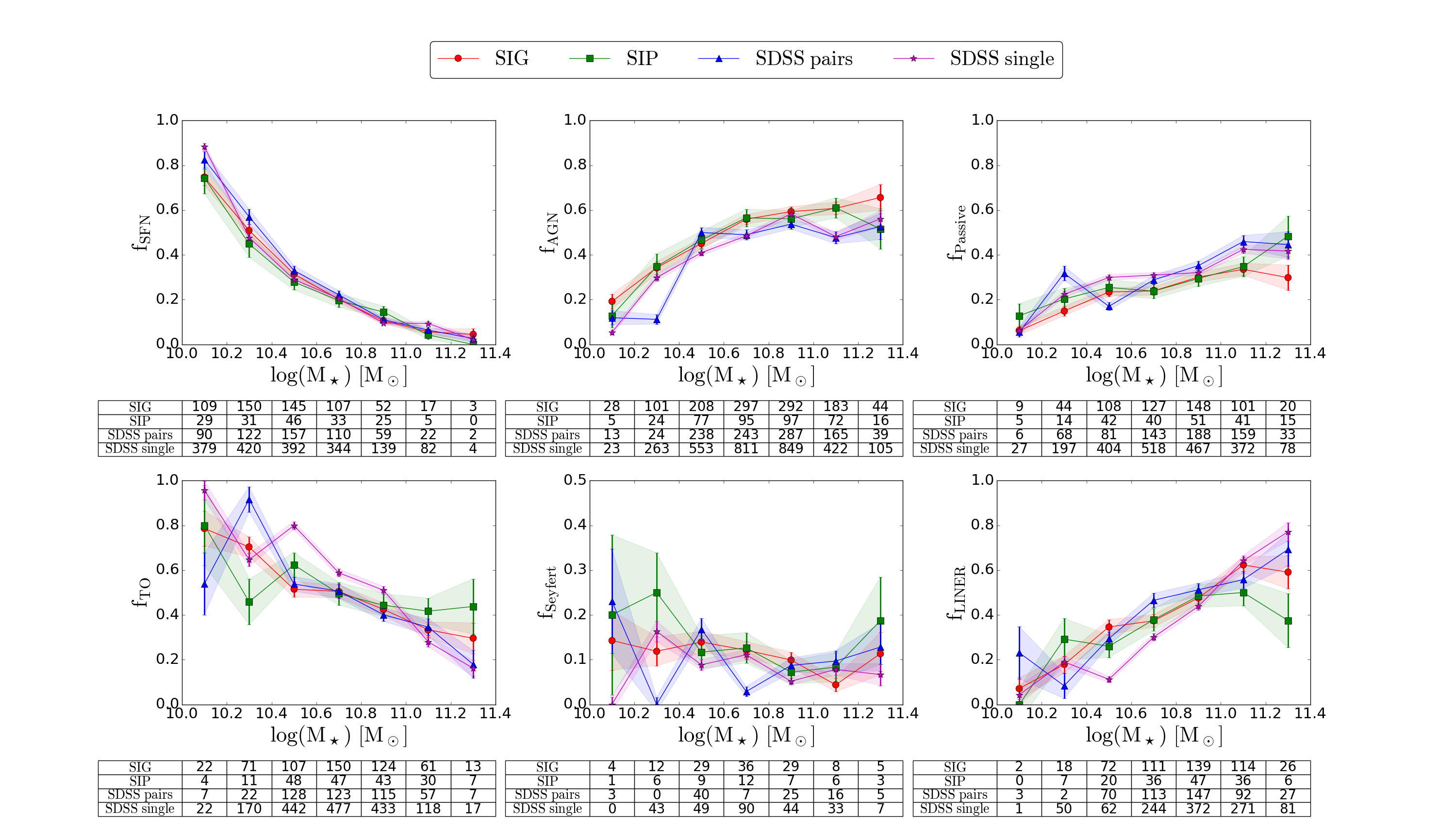}
\caption{{\it Upper panels:} Fraction of SFN (left panel), optical AGN (middle panel), and passive galaxies (right panel) with respect to stellar mass. The fraction in SIG ($N_T = 2299$) and SIP ($N_T$ = 764) galaxies is depicted by red circles and green squares respectively. For the control samples, blue triangles correspond to the fraction in SDSS pairs ($N_T$ = 2251) and magenta stars for SDSS single galaxies ($N_T$ = 6867). The number of galaxies in each stellar mass bin is shown in tables for each sample at the bottom of each panel. Error bars are given by considering binomial distribution. {\it Lower panels:} Fraction of each AGN subtype with respect to stellar mass for SIG ($N_T$ = 1153) and SDSS single ($N_T$ = 3027) galaxies, and central galaxies in SIP ($N_T$ = 387) and SDSS pairs ($N_T$ = 1009). The fraction of LINER (left panel), Seyfert (middle panel), and TO (right panel) galaxies is shown in tables for each sample at the bottom of each panel.}
\label{Fig:f_agn}
\end{figure*}  

\section{Results} \label{Sec:res}

\subsection{AGN prevalence}

We aim to study the prevalence of AGN in the four samples previously selected. To do this we compare the relative fraction of each type of AGN in isolated galaxies and in physically bound isolated pairs, with respect to the ones found in the control samples. Due to the strong dependence of the prevalence of AGN with mass of the host galaxy, both in optical \citep{2003MNRAS.346.1055K} and radio \citep{2005MNRAS.362...25B}, we fix a stellar mass range in each step of the study. We considered galaxies in each sample with stellar masses within $10.0~\leq~\rm{log}(M_\star)~\leq~11.4~[M_\odot]$, as explained in Sect.~\ref{Sec:data}, and we divided our studies in different stellar mass bins.

\subsubsection{Optical AGN}

The relative fraction of optical nuclear activity (SFN, optical AGN, and passive galaxies) for each sample is shown in the upper panels in Fig.~\ref{Fig:f_agn}. We considered seven stellar mass bins to observe the possible trends for each sample. Note that we take care of having a significant number of galaxies in each bin. Error bars are given by considering binomial distribution\footnote{$e_{f}~=~\sqrt{\frac{f(1-f)}{N_T}}$, where $f$ is the relative fraction for a total number of $N_T$ galaxies in each sample.}. 

The fraction of passive galaxies (right panel) increases with stellar mass, while the general trend for SFN galaxies (middle panel) is to decrease with stellar mass. The general trend for the fraction of optical AGN up to $\rm{log}(M_\star)~\lesssim~11.0~[M_\odot]$ is to increase with stellar mass (left panel). We only find significant differences between samples for massive galaxies. The fraction of optical AGN is still increasing at higher masses for isolated galaxies while it starts to decrease in the remaining samples. Accordingly, the fraction of passive isolated galaxies is lower at higher masses.

Lower panels in Fig.~\ref{Fig:f_agn} shows the relative fraction of each optical AGN type (LINER, Seyfert, and TO). When considering optical AGN subtypes, we find that the prevalence of LINER follows the general trend of passive galaxies. On the other hand, the fraction of TOs follows the general trend observed for star-forming galaxies. Again, there is a break point at $\rm{log}(M_\star)~\lesssim~11.0~[M_\odot]$ where we start to observe significant differences between the samples. At high stellar masses there is a higher fraction of TOs and a lower fraction of LINERs SIP galaxies.

\subsubsection{Radio AGN}

\begin{figure}
\centering
\includegraphics[width=\columnwidth]{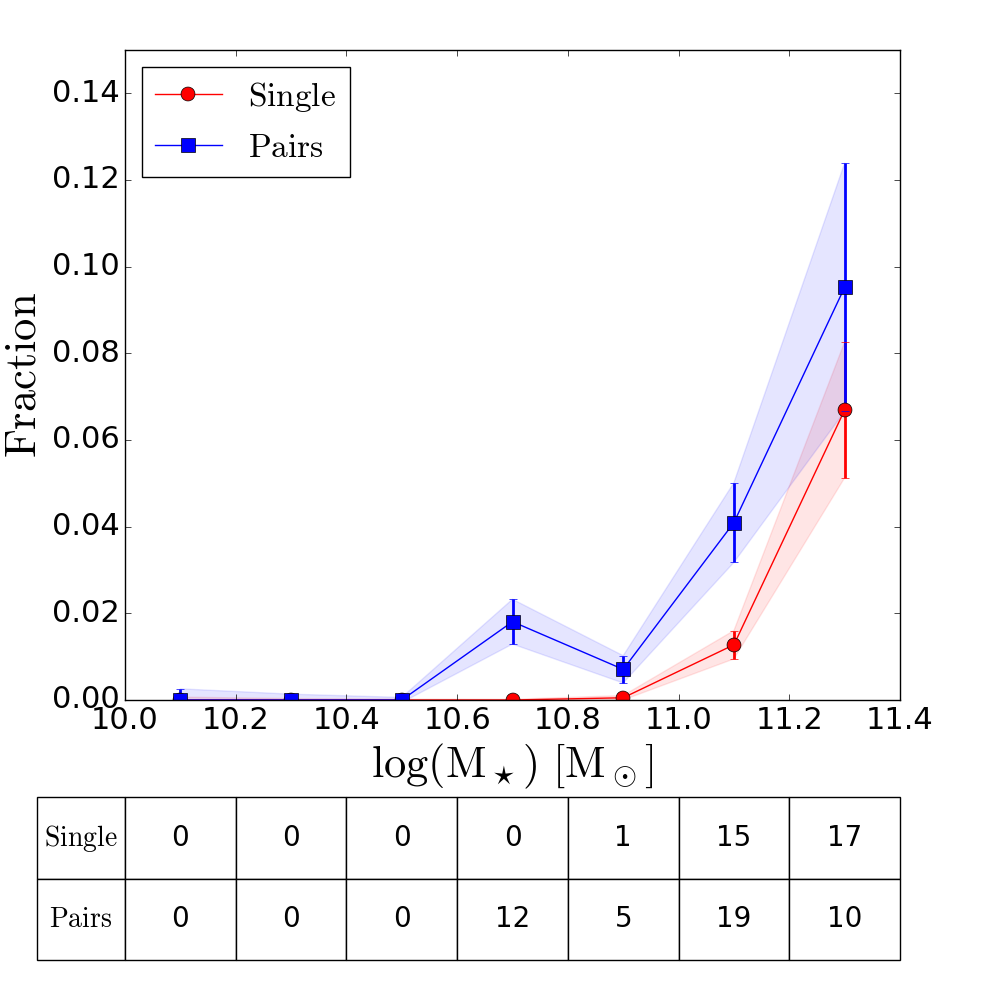}
\caption{Fraction of radio AGN (LERG) with respect to stellar mass. Red circles represent the join sample of SIG and SDSS single galaxies ($N_T$ = 33), and blue squares represent the sample of galaxies in pairs, composed of SIP and SDSS pairs ($N_T$ = 45). The number of galaxies in each stellar mass bin is shown in the table at the bottom. Error bars are given considering binomial distribution.}
\label{Fig:f_agn_radio}
\end{figure}

Given that we have a low number of radio AGN galaxies in each sample (see last row in Table~\ref{tab:samples}), we compare the activity between systems composed of one galaxy (SIP and SDSS single galaxies) and galaxy pairs (SIP and SDSS pairs). The fraction of radio AGN is shown in Fig.~\ref{Fig:f_agn_radio}. The fraction of radio AGN increases steeply with the stellar mass. The prevalence of radio AGN in pairs is significantly higher than in single galaxies for the most massive galaxies.

It would be possible to lower the radio detection limit to study the radio nuclear activity for a wider range of stellar masses \citep{2005MNRAS.362...25B}. The downside is that the closer the Universe less volume and less of each type galaxies, therefore the number of galaxies would be not statistically significant. 

\subsection{Influence of the environment}

After studying the prevalence of optical nuclear activity as a function of mass, we investigate its relation with the local and large-scale galactic environments. We take into account the effect of the mass dividing the samples into three stellar mass bins: low-mass galaxies ($10.0~\leq~\rm{log}(M_\star)~<~10.5~[M_\odot]$), intermediate-mass galaxies ($10.5~\leq~\rm{log}(M_\star)~<~11.0~[M_\odot]$) and high-mass galaxies ($11.0~\leq~\rm{log}(M_\star)~\leq~11.4~[M_\odot]$). As introduced in Sect.~\ref{Sec:data}, we use $Q_{\rm{pair}}$ to quantify the influence of the companion, and $Q_{\rm LSS}$ to quantify the effect of the LSS. Given the low statistics for radio AGN and AGN subtypes, we perform this part of the study on the fraction of optical AGN, SFN, and passive galaxies in each sample. 

\subsubsection{Dependence on the local environment}

According to \citet{2014A&A...564A..94A,2015A&A...578A.110A}, about 99\% of the total tidal strength is due to the effect of the physically bound companions. Therefore, to investigate the dependence of nuclear activity with the local environment we restrict our study to the SIP and the SDSS pairs samples. 

The fraction of nuclear activity, segregated in stellar mass bins, with respect to the $Q_{\rm pair}$ for central galaxy in the SIP and SDSS pairs samples, is shown in the left and right columns in the Fig.~\ref{Fig:f_agn_local}, respectively. Higher values of the $Q_{\rm{pair}}$ correspond to a stronger interaction between the two galaxies in the pair. In general we do not see any trend in the fraction of SFN, AGN, or passive galaxies with the local environment. Moreover, there are no significant differences between the SIP and SDSS samples in the area with common values of $Q_{\rm pair}$. We further explore the dependence on the local environment and discuss these results in Sect.~\ref{Sec:dis}.

\subsubsection{Dependence on the LSS environment}

As introduced in Sect.~\ref{Sec:data}, we use the $Q_{\rm LSS}$ to explore the effect of the large-scale environment on the fraction of nuclear activity. We do not find significant differences in the trends for the SIG and SIP samples, as well as for the control samples. Henceforth, for a clearer analysis of the results, we focus on the comparison between SIG and SDSS single galaxies to explore the effect of the large-scale environment.

Similarly to Fig.~\ref{Fig:f_agn_local}, Fig.~\ref{Fig:f_agn_lss} shows the fraction of optical nuclear activity, segregated in stellar mass bins, with respect to $Q_{\rm LSS}$, in this case for the SIG and SDSS single galaxies. In general, there is no clear dependence on the large-scale environment in the nuclear activity for SDSS single galaxies. However there is a strong effect in isolated galaxies, with different trends depending on stellar mass. We discuss these trends in more details in Sect.~\ref{Sec:dis}.

\section{Discussion}  \label{Sec:dis}

\subsection{AGN prevalence}  \label{Sec:massdepence}

We study the optical and radio AGN prevalence in each sample to explore the different mechanisms triggering nuclear activity. In particular, we will relate the obtained results to the different sources of gas that can fuel AGN. Depending on the nature (environment) of the studied samples, these mechanisms could be: 1) the internal and slow mass loss caused by the central BH (internal secular evolution), 2) the external and slow prolonged gas infall or the galaxy harassment (environmental secular evolution), or 3) the external and fast ram-pressure striping or galaxy mergers.

\subsubsection{Optical AGN}

The general trend of optical AGN and passive galaxies is to increase with stellar mass, while the fraction of star forming galaxies decreases from low-mass to massive galaxies, as it is shown in the upper panels in Fig.~\ref{Fig:f_agn}. These monotonic trends with stellar mass are expected because of the 'downsizing' effect.

We find that, in general, there is no difference in the fraction of optical AGN for the four samples. This result points out that optical AGN is independent of the local environment, and is therefore in agreement with the general consensus that secular evolution is a main mechanism triggering optical AGN in the Local Universe \citep{2011RMxAA..47..361C,2012A&A...545A..15S,2013MNRAS.430..638S,2013MNRAS.434..336H,2015MNRAS.447..110S,2015MNRAS.447.2209P,2016MNRAS.tmp..262H}. As a consequence, the prevalence of optical nuclear activity is independent to the addition of one single companion: the central galaxy in an isolated pair has no difference with an isolated galaxy. We discuss the discrepancies with other studies when considering the effects of the galaxy pair in Sect.~\ref{Sec:envdepencelocal}.

However, we observe significant differences for higher stellar mass bins ($M_\star~>~10^{11.0}\,M_\odot$). This effect is in agreement with \citet{2015MNRAS.451.1482M}, which claim that the environmental influence is notable in the highest mass galaxies. For massive galaxies, the fraction of optical AGN in the SIG and SIP samples is slightly higher than in control samples. In particular, the fraction of optical AGN for massive isolated galaxies is higher than in any other sample, in discrepancy with \citet{2015MNRAS.451.1482M}, where the fraction of AGN in pairs is only a little higher than in isolated galaxies. 

According to the redshift distribution of the samples (lower panel of Fig.~\ref{Fig:hist_mass}), we consider that the trends observed in the fraction of AGN subtypes (see lower panels in Fig.~\ref{Fig:f_agn}) are real. Otherwise the trends would be roughly constant in case that a high fraction of low luminosity AGN at low redshift were misclassified as LINERs. These observed trends suggest that low-mass AGNs are dominated by TOs, while high-mass AGNs are dominated by LINERs. 

\citetalias{2013MNRAS.430..638S} interpreted the time sequence from interaction inducing star formation to passive galaxies \citep{2008MNRAS.385.1915L,2010MNRAS.405..933W} in order from TO, then Seyfert, and then LINER. Even if our statistic for Seyfert-like galaxies is small, we can interpret our result in light of this sequence. According to our results, the transition from TO to LINER is slower in isolated pairs. Unfortunately we do not have a statistically meaningful number of SIP pairs ($N_T~=~191,44,152$ LINERs, Seyferts, and TOs, respectively) to explore this result as a function of the $Q_{\rm pair}$.

These results suggest that the black holes of massive ($M_\star~>~10^{11.0}\,M_\odot$) isolated galaxies are still growing while similar mass isolated pairs, SDSS pairs, and SDSS single galaxies have quenched their activity. This value is similar to the transition mass between hot and cold modes of gas accretion in simulations by \citet{2009MNRAS.395..160K}. We can therefore conclude that cold gas accretion by secular evolution is sufficient to explain the optical nuclear activity for more massive galaxies. As suggested by \citetalias{2013MNRAS.430..638S}, the decrease of the prevalence of optical AGN and LINER for massive galaxies in denser environments can be explained by the striping of cold gas and its warming.

\subsubsection{Radio AGN}

In the exploration of the possible sources of gas fuelling radio AGN and its connection to the environment, it is crucial to discriminate between the different radio AGN modes \citep{2006MNRAS.365...11C,2007MNRAS.376.1849H,2008A&A...490..893T}. In fact, \citetalias{2013MNRAS.430..638S} found opposite trends in HERG and LERG radio AGNs with respect to the local density, when LERGs show a clear increase with density. As it is explained in Sect.~\ref{Sec:AGN}, in this study we focus on LERG radio AGNs. It is important to note that the nature of radio AGN is also sensitive to redshift evolution \citep{2014ApJ...797...26K,2016MNRAS.457..629C,2016MNRAS.456..431M}. Since the galaxies with \citetalias{2013MNRAS.430..638S} classification are restricted to the narrow redshift range $0.03 \leq z \leq 0.08$ and the samples follow a similar redshift distribution, we do not expect a bias in our comparison caused by the possible redshift evolution. 

Generally, radio nuclear activity is strongly related to the stellar mass and the density of galaxies \citepalias{2013MNRAS.430..638S}. Hence, we do not expect to find a high fraction of radio AGN in low density clusters. This dependence with the galaxy density (large-scale environment) is even observed in isolated galaxies. \citet{2008A&A...486...73S,2012A&A...545A..15S} do not find any high luminosity radio AGN in isolated galaxies in the AMIGA \citep[\textbf{A}nalysis of the interstellar \textbf{M}edium of \textbf{I}solated \textbf{GA}laxies,][]{2005A&A...436..443V} sample. In fact, the fraction of radio AGN in isolated galaxies is smaller than expected for galaxies with same stellar mass.

A recent study of radio nuclear activity for local galaxies \citep{2015MNRAS.452..774K} discarded AGN to be triggered by internal mass loss (secular processes). In relation to the large-scale environment, they also dismissed that radio AGN are fuelled by cluster-scale cooling flows since their radio detections preferentially lie outside clusters. They therefore conclude that the most important trigger for 'cold-mode' radio AGN is galaxy merging. In this regard, \citet{2012MNRAS.419..687R} and \citet{2015ApJ...806..147C} found strong evidence that mergers are the triggering mechanism for the radio-loud AGN phenomenon. Note that the radio-loud AGNs studied are distinct from the LERGs.

The fraction of radio AGN for single galaxies ($N_T$\,=\,33) and galaxies in pairs ($N_T$\,=\,45) in our study is shown in Fig.~\ref{Fig:f_agn_radio}. The dependence of radio AGN with stellar mass is so strong that, even if the number of radio AGN detections in our samples is small, we can see significant differences with the addition of one single companion. These differences reflect that not only the LSS affects the radio AGN activity, we also found evidence that radio AGN is affected by the local environment. This result is in agreement with \citet{2014ApJ...785...66P}, whose find that radio AGN (LERGs) tend to be located in dense environments. They also claim that they are fuelled by the accretion of small quantities of hot halo gas. Therefore, while the effect of interactions is minimal in triggering optical AGN activity, it seems to have a strong connection with radio AGN activity. Unfortunately, we do not have a statistically significant sample to further explore the dependence on the local and large-scale environments as for optical AGN in this study.

Note here that the differences between the two samples could be reduced if we consider halo masses instead of stellar masses. In fact, \citet{2015MNRAS.451L..35E} found that radio AGN (LERGs) are not fuelled by mergers, since they do not find an excess on the fraction of radio AGN (LERGs) when matching control samples in halo mass or D4000. Moreover, the results of Fig.~\ref{Fig:f_agn_radio} agree with the results of \citet{2015MNRAS.451L..35E}, whose find an excess of LERGs in pairs if only stellar-mass and redshift are matched.

\subsection{Dependence on the environment} \label{Sec:envdepence}

There has been some disagreement in the literature regarding the connection between environment and nuclear activity. In this context, we explore the effect of local and large-scale environments on the AGN activity in isolated galaxies and physically bound isolated pairs, selected under a three-dimensional isolation definition. To investigate this, we computed the tidal strengths $Q_{\rm{pair}}$ and $Q_{\rm{LSS}}$ as explained in Sect.~\ref{Sec:data}. The different nature of the different environment definitions for isolated systems and the control samples are shown in Fig.~\ref{Fig:Qsamples}. By definition, control samples dominate the range of higher values of the $Q_{\rm{LSS}}$, while the systems of galaxies in pairs lie in the area of higher $Q_{\rm{pair}}$. 

\begin{figure*}
\centering
\includegraphics[width=2.\columnwidth]{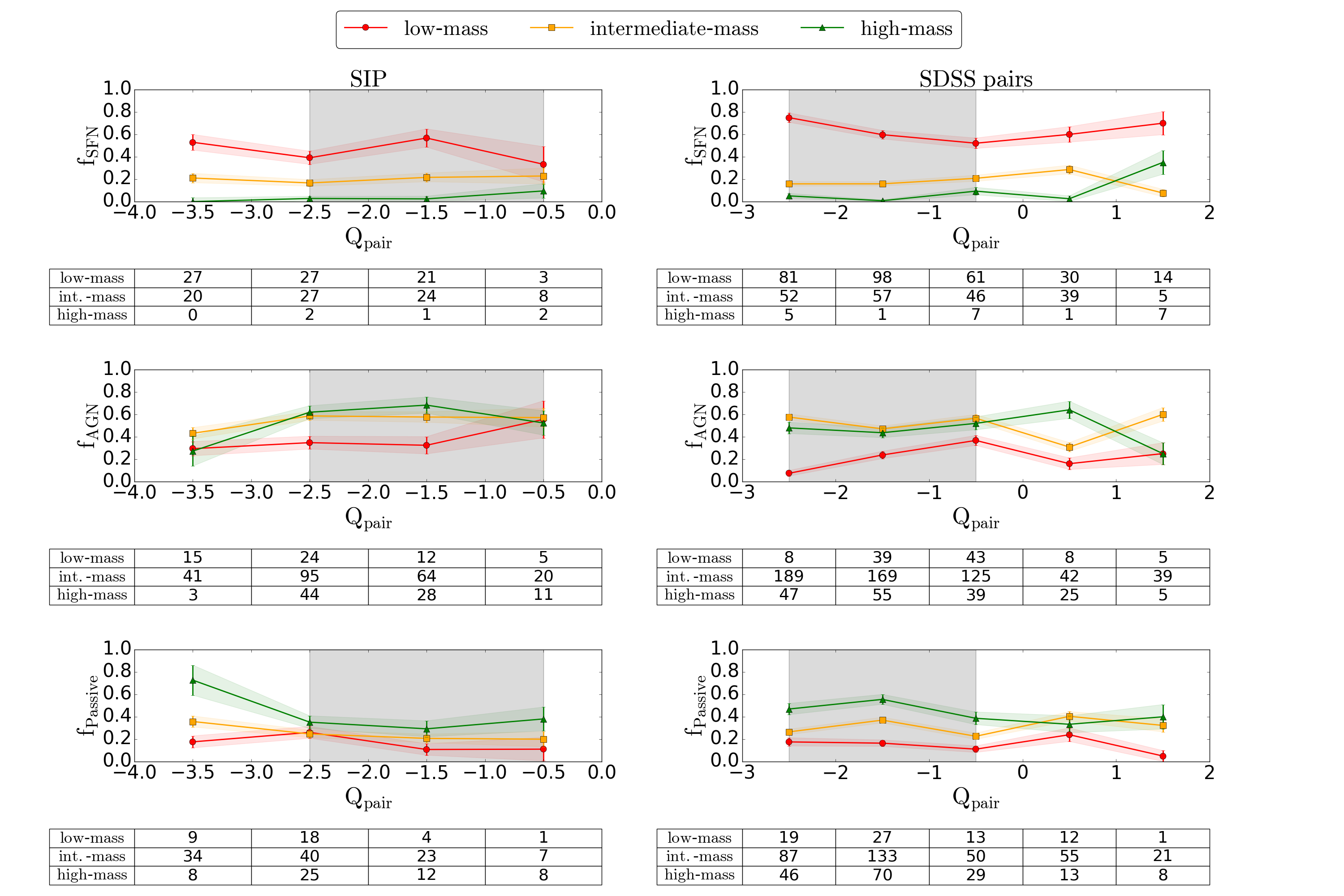}
\caption{Fraction of optical nuclear activity with respect to the $Q_{\rm pair}$ environmental parameter. Low-mass galaxies ($10.0~\leq~\rm{log}(M_\star)~<~10.5~[M_\odot]$) are represented by red circles, intermediate-mass galaxies ($10.5~\leq~\rm{log}(M_\star)~<~11.0~[M_\odot]$) are represented by orange squares, and high-mass galaxies ($11.0~\leq~\rm{log}(M_\star)~\leq~11.4~[M_\odot]$) are represented by green triangles. The fraction of SFN, optical AGN, and passive SIP galaxies ($N_T$ = 169, 387, and 208, respectively) is represented from top to bottom in the left panels, and for SDSS pairs ($N_T$ = 564, 1009, and 678, respectively) in the right panels. The number of galaxies in each $Q_{\rm pair}$ bin is shown in tables for each sample at the bottom of each panel. The dashed area in the figures corresponds to the range with common values of $Q_{\rm{pair}}$ between the two samples, from -2.5 to -0.5. Error bars are given considering binomial distribution.}
\label{Fig:f_agn_local}
\end{figure*}

\subsubsection{Dependence on the local environment}  \label{Sec:envdepencelocal}

The values of $Q_{\rm pair}$ for SDSS pairs are mainly larger than the values for SIP galaxies, as it is shown in Fig.~\ref{Fig:Qsamples}. This is expected since there are more SDSS pairs at smaller projected distances and higher mass ratio than in isolated pairs. Nevertheless, there is an area with common values of the $Q_{\rm pair}$ between -2.5 to -0.5 (see Fig.~\ref{Fig:f_agn_local}). We do not observe relevant differences in this common area. In general, we do not observe any significant trend as a function of $Q_{\rm pair}$ (see Fig.~\ref{Fig:f_agn_local}), neither as a function of $\Delta Q$, which is defined as $Q_{\rm pair}~-~Q_{\rm LSS}$ and quantifies the extent that a given galaxy is dominated by its closest neighbour or its large-scale environment.

On the contrary, \citet{2011MNRAS.418.2043E} found an increase in the AGN fraction in most tight pairs and stated that optical AGN might be triggered by close interactions. A possible explanation for the discrepancy is the fairly wide projected sepatarions of the SDSS pairs sample ($d~\simeq~200$\,kpc on average) in comparison to the AGN excess seen in \citet{2011MNRAS.418.2043E} at distances $d~\lesssim~50$\,kpc. To check this we further explore the fraction of optical AGN as a function of the projected separation and the mass ratio between the two galaxies in pairs. Even if we do not have a large statistical sample of isolated pairs with projected separation smaller than 50\,kpc ($N_T$~=~50), we have a large number ($N_T$~=~435) of close SDSS pairs to compare with the results of \citet{2011MNRAS.418.2043E}. We also do not observe any trend and any relevant difference between the SIP or the SDSS pairs samples with projected separation. This discrepancy might then come from the different definitions of the local environment. For the purpose of the present study, galaxy pairs are located in low density environments. By definition, the SIP sample is very well isolated from the large-scale environment, with the first nearest neighbour at projected distances larger than 1\,Mpc. In the case of SDSS pairs, we selected groups in \citet{2007ApJ...671..153Y} with two galaxies within the same dark matter halo to avoid the case where the same central galaxy can be included in two or three different pairs. We therefore confirm our previous result that local environment has not a principal role in triggering optical AGN. 

The parameters described in Eqs.~\ref{Eq:QLSS} and \ref{Eq:Qlocal} quantify the tidal effects onto the central (brightest) galaxy in galaxy pairs. In this sense, high values of these parameters, in relation to the AGN fraction, are mainly associated with the gas stripping or the shape distortion of the central galaxy. Nevertheless, AGN properties are more associated with the gas accretion of the central galaxy. In this sense, an alternative definition of the $Q_{\rm pair}$ ,i.e. $Q_{\rm pair,B} \equiv {\rm log} \left(\frac{M_{A}}{M_{B}} \left(\frac{D_{B}}{d_{AB}}\right)^3\right)$, is more closely related to trigger the gas accretion to the central galaxy. To better identify the different contributions it is worth to further explore the tidal strength affecting the faintest galaxy in the galaxy pairs. However, this should be carried out in a separated study since only 45\% of the SIP has \citetalias{2013MNRAS.430..638S} classification for the two members in the mass range considered in the present study \citep[the typical mass ratio is $\sim1/100$][]{2015A&A...578A.110A}.

\begin{figure*}
\centering
\includegraphics[width=2.\columnwidth]{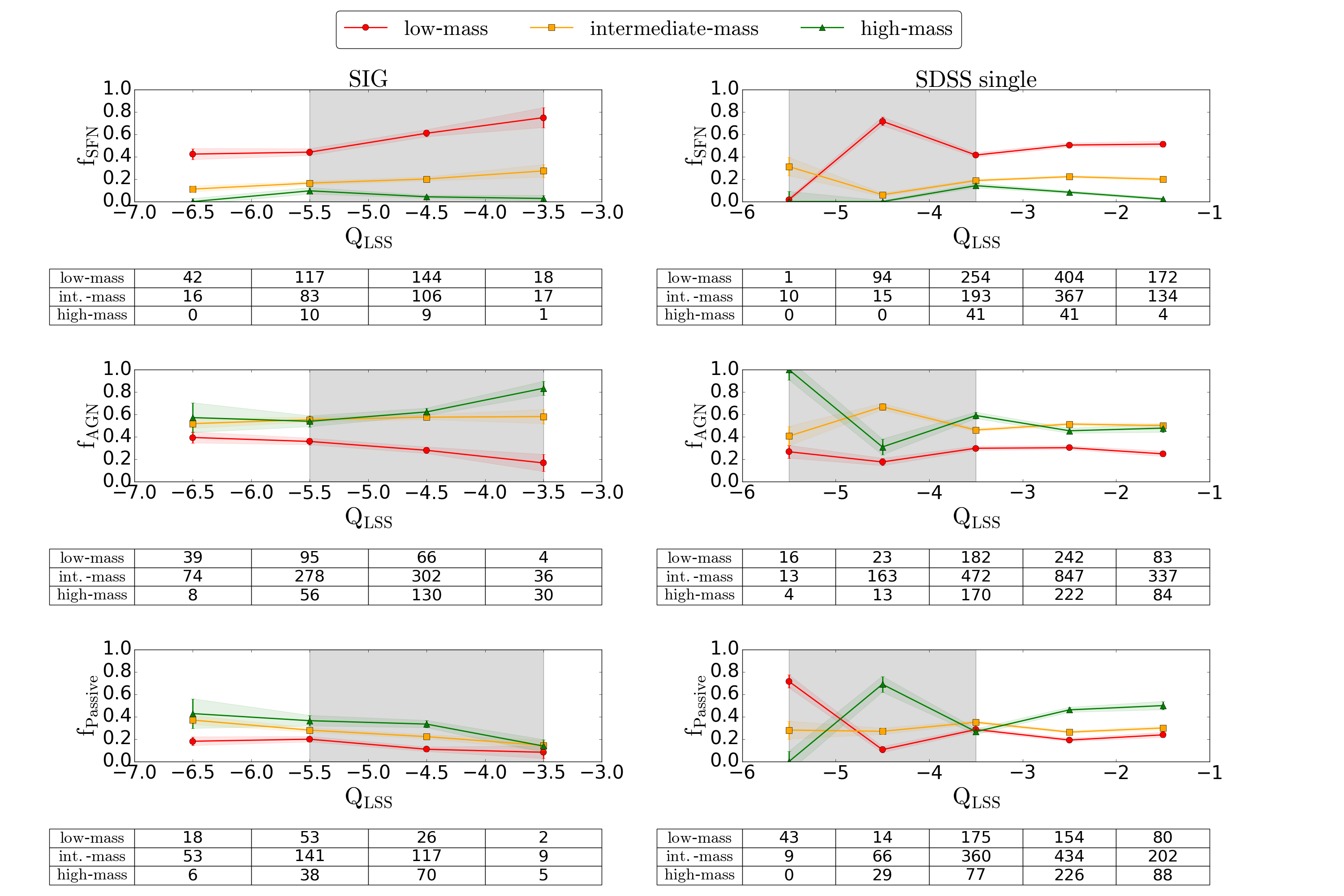}
\caption{Fraction of optical nuclear activity with respect to the $Q_{\rm LSS}$ environmental parameter. Low-mass galaxies ($10.0~\leq~\rm{log}(M_\star)~<~10.5~[M_\odot]$) are represented by red circles, intermediate-mass galaxies ($10.5~\leq~\rm{log}(M_\star)~<~11.0~[M_\odot]$) are represented by orange squares, and high-mass galaxies ($11.0~\leq~\rm{log}(M_\star)~\leq~11.4~[M_\odot]$) are represented by green triangles. The fraction of SFN, optical AGN, and passive SIG galaxies ($N_T$ = 587, 1153, and 559, respectively) is represented from top to bottom in the left panels, and for SDSS single galaxies ($N_T$ = 1777, 3027, and 2063, respectively) in the right panels. The number of galaxies in each $Q_{\rm LSS}$ bin is shown in tables for each sample at the bottom of each panel. The dashed area in the figures corresponds to the range with common values of $Q_{\rm LSS}$ between the two samples, from -5.5 to -3.5. Error bars are given considering binomial distribution.}
\label{Fig:f_agn_lss}
\end{figure*}

\subsubsection{Dependence on the LSS environment} \label{Sec:envdepencelss}

In contrast to SDSS single galaxies, we find a strong dependence of the LSS on the optical nuclear activity and star formation in isolated galaxies. Moreover, the observed trends are different depending on galaxy mass (see Fig.~\ref{Fig:f_agn_lss}). 

According to \citet{2015A&A...578A.110A}, SIG galaxies mainly belong to the outer parts of filaments, walls, and clusters, and generally differ from the void population of galaxies. In fact, only one third of SIG galaxies are located in voids. Using the code for data visualisation LSSGALPY\footnote{\texttt{https://github.com/margudo/LSSGALPY}} \citep{2015ascl.soft05012A}, we checked that galaxies with low values of $Q_{\rm{LSS}}$ are mainly located in void regions, and galaxies with higher $Q_{\rm{LSS}}$ are more related with denser structures, as filaments or walls. See Appendix~\ref{Sec:App1} for more details.

This means that the differences that we found between SIG galaxies and SDSS singles, with respect to high values of the $Q_{\rm{LSS}}$, are directly related to the location of the galaxies in the outskirts or inside clusters, respectively. According to this, the general trend for the fraction of passive isolated galaxies is to decrease from voids to denser regions (e. g. clusters, filaments). However, the fraction is smaller than that of similar mass SDSS single galaxies located in the same regions. 

The fraction of optical AGN for high-mass SIG galaxies increases with denser large-scale environment (massive SIG galaxies have more accretion time and therefore they have more gas). We interpret this result as the central black hole in massive isolated galaxies being fuelled by cold gas from the LSS. We would not see an increment in SDSS single galaxies because of the warming of the cold gas in denser environments, where the effect of the stripping of diffuse gas or strangulation, ram-pressure stripping, and galaxy harassment is present. \citep{2010MNRAS.404.1231V,2015Natur.521..192P}. 

On the contrary, the fraction of optical AGN for low-mass SIG galaxies decreases from voids to denser regions (e.g. clusters, filaments). In comparison to the trend for passive galaxies, this is directly translated into an important increment of star-forming SIG galaxies in the outskirts of clusters. Given that the fraction of star-forming SIG galaxies with low values of $Q_{\rm{LSS}}$ is smaller, our results contrast to those of \citet{2015ApJ...810..165L}. They found that the fraction of star-forming galaxies in voids is significantly higher than that in walls. Note that only 14\% of void galaxies in \citet{2012MNRAS.421..926P} are found in the SIG sample \citep{2015A&A...578A.110A}, therefore void galaxies span a large range of local environments. In the future, we will explore these differences further by considering the possible effect of morphology or stellar populations.

We previously discussed the fact that the fraction of optical AGN in the SIG sample continues increasing, while peaks at $M_\star~\simeq~10^{8.0}\,M_\odot$ for the other samples (see Fig.~\ref{Fig:f_agn}). From Fig.~\ref{Fig:f_agn_lss}, we see that this continue increase mainly occurs at high $Q_{\rm{LSS}}$. It is interesting that, on the contrary, the AGN fraction of the low-mass SIG galaxies even decreases. Such a different mass dependent worth a further detailed study. 

The different mass- and environment-dependence behaviours between isolated and control samples suggest that a halo/mass or a simple $Q_{\rm{LSS}}$ parameter is not enough to characterise the environmental effects of galaxies in more complicated (small-scale) environments.

\section{Summary and conclusions} \label{Sec:con}

In this work we study the effect of the environment on the fraction of optical and radio nuclear activity. In particular, we investigate the effect of both, local and large-scale environments on nuclear activity and star formation, for the first time using three-dimensional isolated galaxies and physically bound isolated pairs (SIG and SIP galaxies, respectively). Besides, using the tidal strength parameters $Q_{\rm pair}$ and $Q_{\rm LSS}$ we are able to  quantify separately the effect of one-on-one interactions from the effect of the large-scale environment where the galaxy resides. Control samples of single galaxies and pairs selected from the SDSS were used for comparison. 

Our main conclusions are the following:

\begin{enumerate}
 \item The prevalence of optical AGN is found to be independent to the addition of one single companion (see Fig.~\ref{Fig:f_agn}).  
 \item For massive galaxies, the fraction of optical AGN in isolated galaxies and isolated pairs is slightly higher than in the control samples (see the central upper panel in Fig.~\ref{Fig:f_agn}). Moreover, the fraction of massive passive isolated galaxies is smaller than in any other sample (see the right upper panel in Fig.~\ref{Fig:f_agn}). These results suggest that the black holes of massive ($\rm{log}(M_\star)\,>\,11.0\,[M_\odot]$) isolated galaxies are still growing while similar mass isolated pairs, SDSS pairs, and SDSS single galaxies have already quenched their activity.
 \item Local environment has not a principal role in triggering optical AGN (see Fig.~\ref{Fig:f_agn_local}). Cold gas accretion by secular evolution is sufficient to explain the optical nuclear activity for more massive galaxies (see the middle left panel in Fig.~\ref{Fig:f_agn_lss}).
 \item In contrast to local environment, we find a strong dependence of the optical nuclear activity and star formation on the LSS for isolated systems (see Fig.~\ref{Fig:f_agn_lss}). In particular, the fraction of AGN for high-mass SIG galaxies increases with denser large-scale environment. We interpret this result as the central black hole in massive isolated galaxies being fuelled by cold gas from the LSS. 
 \item Regarding to radio nuclear activity (LERG), we find that not only the galaxy mass and large-scale environment affects it, but that radio AGN are also strongly affected by the local environment (see Fig.~\ref{Fig:f_agn_radio}).
\end{enumerate}

Optical AGN is related to cold gas accretion, while radio AGN is related to hot gas accretion. In this context, there is more cold gas, fuelling the central optical AGN, in isolated systems. Overall, our results are in agreement with a scenario where cold gas accretion by secular evolution is the main driver for optical AGN, while hot gas accretion and one-on-one interactions are the main drivers of radio AGN activity.

\section*{Acknowledgments}

The authors acknowledge the anonymous referee for his/her very detailed and useful report, which helped to clarify and improve the quality of this work.

MAF is grateful for financial support from PIFI (funded by Chinese Academy of Sciences President's International Fellowship Initiative) Grant No. 2015PM056 and from CONICYT FONDECYT project No. 3160304. This work was partly supported by the Strategic Priority Research Program ''The Emergence of Cosmological Structures'' of the Chinese Academy of Sciences (CAS; grant XDB09030200), the National Natural Science Foundation of China (NSFC) with the Project Number of  11573050 and 11433003, and the ''973 Program'' 2014 CB845705. This work was partially supported by MInisterio de Economia y Competividad and by FEDER (Fondo Europeo de Desarrollo Regional) via grants AYA2011-24728, AYA2013-47742-C04-01, AYA2014-53506-P, and from the ''Junta de Andalucía'' local government through the FQM-108 project.

This research made use of \textsc{astropy}, a community-developed core \textsc{python} ({\tt http://www.python.org}) package for Astronomy \citep{2013A&A...558A..33A}; \textsc{ipython} \citep{PER-GRA:2007}; \textsc{matplotlib} \citep{Hunter:2007}; \textsc{numpy} \citep{:/content/aip/journal/cise/13/2/10.1109/MCSE.2011.37}; \textsc{scipy} \citep{citescipy}; and \textsc{topcat} \citep{2005ASPC..347...29T}.

Funding for SDSS-III has been provided by the Alfred P. Sloan Foundation, the Participating Institutions, the National Science Foundation, and the U.S. Department of Energy Office of Science. The SDSS-III web site is http://www.sdss3.org/. 

\bibliography{astroph}

\begin{thebibliography}{86}
\expandafter\ifx\csname natexlab\endcsname\relax\def\natexlab#1{#1}\fi

\bibitem[{{Abazajian} {et~al.}(2009){Abazajian}, {Adelman-McCarthy},
  {Ag{\"u}eros}, {Allam}, {Allende Prieto}, {An}, {Anderson}, {Anderson},
  {Annis}, {Bahcall}, \& et~al. {et al.}}]{2009ApJS..182..543A}
{Abazajian}, K.~N., {Adelman-McCarthy}, J.~K., {Ag{\"u}eros}, M.~A., {et~al.}
  2009, \apjs, 182, 543

\bibitem[{{Abell}(1958)}]{1958ApJS....3..211A}
{Abell}, G.~O. 1958, \apjs, 3, 211

\bibitem[{{Ahn} {et~al.}(2014){Ahn}, {Alexandroff}, {Allende Prieto}, {Anders},
  {Anderson}, {Anderton}, {Andrews}, {Aubourg}, {Bailey}, {Bastien}, \&
  et~al.}]{2014ApJS..211...17A}
{Ahn}, C.~P., {Alexandroff}, R., {Allende Prieto}, C., {et~al.} 2014, \apjs,
  211, 17

\bibitem[{{Alam} {et~al.}(2015){Alam}, {Albareti}, {Allende Prieto}, {Anders},
  {Anderson}, {Anderton}, {Andrews}, {Armengaud}, {Aubourg}, {Bailey}, \&
  et~al.}]{2015ApJS..219...12A}
{Alam}, S., {Albareti}, F.~D., {Allende Prieto}, C., {et~al.} 2015, \apjs, 219,
  12

\bibitem[{{Alexander} \& {Hickox}(2012)}]{2012NewAR..56...93A}
{Alexander}, D.~M. \& {Hickox}, R.~C. 2012, \nar, 56, 93

\bibitem[{{Argudo-Fern{\'a}ndez}
  {et~al.}(2015{\natexlab{a}}){Argudo-Fern{\'a}ndez}, {Duarte Puertas},
  {Verley}, {Sabater}, \& {Ruiz}}]{2015ascl.soft05012A}
{Argudo-Fern{\'a}ndez}, M., {Duarte Puertas}, S., {Verley}, S., {Sabater}, J.,
  \& {Ruiz}, J.~E. 2015{\natexlab{a}}, {LSSGALPY: Visualization of the
  large-scale environment around galaxies on the 3D space}, Astrophysics Source
  Code Library, record ascl:1505.012

\bibitem[{{Argudo-Fern{\'a}ndez}
  {et~al.}(2015{\natexlab{b}}){Argudo-Fern{\'a}ndez}, {Verley}, {Bergond},
  {Duarte Puertas}, {Ramos Carmona}, {Sabater}, {Fern{\'a}ndez Lorenzo},
  {Espada}, {Sulentic}, {Ruiz}, \& {Leon}}]{2015A&A...578A.110A}
{Argudo-Fern{\'a}ndez}, M., {Verley}, S., {Bergond}, G., {et~al.}
  2015{\natexlab{b}}, \aap, 578, A110

\bibitem[{{Argudo-Fern{\'a}ndez} {et~al.}(2014){Argudo-Fern{\'a}ndez},
  {Verley}, {Bergond}, {Sulentic}, {Sabater}, {Fern{\'a}ndez Lorenzo},
  {Espada}, {Leon}, {S{\'a}nchez-Exp{\'o}sito}, {Santander-Vela}, \&
  {Verdes-Montenegro}}]{2014A&A...564A..94A}
{Argudo-Fern{\'a}ndez}, M., {Verley}, S., {Bergond}, G., {et~al.} 2014, \aap,
  564, A94

\bibitem[{{Argudo-Fern{\'a}ndez} {et~al.}(2013){Argudo-Fern{\'a}ndez},
  {Verley}, {Bergond}, {Sulentic}, {Sabater}, {Fern{\'a}ndez Lorenzo}, {Leon},
  {Espada}, {Verdes-Montenegro}, {Santander-Vela}, {Ruiz}, \&
  {S{\'a}nchez-Exp{\'o}sito}}]{2013A&A...560A...9A}
{Argudo-Fern{\'a}ndez}, M., {Verley}, S., {Bergond}, G., {et~al.} 2013, \aap,
  560, A9

\bibitem[{{Astropy Collaboration} {et~al.}(2013){Astropy Collaboration},
  {Robitaille}, {Tollerud}, {Greenfield}, {Droettboom}, {Bray}, {Aldcroft},
  {Davis}, {Ginsburg}, {Price-Whelan}, {Kerzendorf}, {Conley}, {Crighton},
  {Barbary}, {Muna}, {Ferguson}, {Grollier}, {Parikh}, {Nair}, {Unther},
  {Deil}, {Woillez}, {Conseil}, {Kramer}, {Turner}, {Singer}, {Fox}, {Weaver},
  {Zabalza}, {Edwards}, {Azalee Bostroem}, {Burke}, {Casey}, {Crawford},
  {Dencheva}, {Ely}, {Jenness}, {Labrie}, {Lim}, {Pierfederici}, {Pontzen},
  {Ptak}, {Refsdal}, {Servillat}, \& {Streicher}}]{2013A&A...558A..33A}
{Astropy Collaboration}, {Robitaille}, T.~P., {Tollerud}, E.~J., {et~al.} 2013,
  \aap, 558, A33

\bibitem[{{Baldry} {et~al.}(2004){Baldry}, {Glazebrook}, {Brinkmann},
  {Ivezi{\'c}}, {Lupton}, {Nichol}, \& {Szalay}}]{2004ApJ...600..681B}
{Baldry}, I.~K., {Glazebrook}, K., {Brinkmann}, J., {et~al.} 2004, \apj, 600,
  681

\bibitem[{{Baldwin} {et~al.}(1981){Baldwin}, {Phillips}, \&
  {Terlevich}}]{1981PASP...93....5B}
{Baldwin}, J.~A., {Phillips}, M.~M., \& {Terlevich}, R. 1981, \pasp, 93, 5

\bibitem[{{Becker} {et~al.}(1995){Becker}, {White}, \&
  {Helfand}}]{1995ApJ...450..559B}
{Becker}, R.~H., {White}, R.~L., \& {Helfand}, D.~J. 1995, \apj, 450, 559

\bibitem[{{Bell} {et~al.}(2003){Bell}, {McIntosh}, {Katz}, \&
  {Weinberg}}]{2003ApJS..149..289B}
{Bell}, E.~F., {McIntosh}, D.~H., {Katz}, N., \& {Weinberg}, M.~D. 2003, \apjs,
  149, 289

\bibitem[{{Best} \& {Heckman}(2012)}]{2012MNRAS.421.1569B}
{Best}, P.~N. \& {Heckman}, T.~M. 2012, \mnras, 421, 1569

\bibitem[{{Best} {et~al.}(2005{\natexlab{a}}){Best}, {Kauffmann}, {Heckman},
  {Brinchmann}, {Charlot}, {Ivezi{\'c}}, \& {White}}]{2005MNRAS.362...25B}
{Best}, P.~N., {Kauffmann}, G., {Heckman}, T.~M., {et~al.} 2005{\natexlab{a}},
  \mnras, 362, 25

\bibitem[{{Best} {et~al.}(2005{\natexlab{b}}){Best}, {Kauffmann}, {Heckman}, \&
  {Ivezi{\'c}}}]{2005MNRAS.362....9B}
{Best}, P.~N., {Kauffmann}, G., {Heckman}, T.~M., \& {Ivezi{\'c}}, {\v Z}.
  2005{\natexlab{b}}, \mnras, 362, 9

\bibitem[{{Blanton} \& {Roweis}(2007)}]{2007AJ....133..734B}
{Blanton}, M.~R. \& {Roweis}, S. 2007, \aj, 133, 734

\bibitem[{{Blanton} {et~al.}(2005){Blanton}, {Schlegel}, {Strauss},
  {Brinkmann}, {Finkbeiner}, {Fukugita}, {Gunn}, {Hogg}, \v{Z}. {Ivezi{\'c}},
  {Knapp}, {Lupton}, {Munn}, {Schneider}, {Tegmark}, \&
  {Zehavi}}]{2005AJ....129.2562B}
{Blanton}, M.~R., {Schlegel}, D.~J., {Strauss}, M.~A., {et~al.} 2005, \aj, 129,
  2562

\bibitem[{{Boselli} \& {Gavazzi}(2014)}]{2014A&ARv..22...74B}
{Boselli}, A. \& {Gavazzi}, G. 2014, \aapr, 22, 74

\bibitem[{{Brinchmann} {et~al.}(2004){Brinchmann}, {Charlot}, {White},
  {Tremonti}, {Kauffmann}, {Heckman}, \& {Brinkmann}}]{2004MNRAS.351.1151B}
{Brinchmann}, J., {Charlot}, S., {White}, S.~D.~M., {et~al.} 2004, \mnras, 351,
  1151

\bibitem[{{Calvi} {et~al.}(2012){Calvi}, {Poggianti}, {Fasano}, \&
  {Vulcani}}]{2012MNRAS.419L..14C}
{Calvi}, R., {Poggianti}, B.~M., {Fasano}, G., \& {Vulcani}, B. 2012, \mnras,
  419, L14

\bibitem[{{Casado} {et~al.}(2015){Casado}, {Ascasibar}, {Gavil{\'a}n},
  {Terlevich}, {Terlevich}, {Hoyos}, \& {D{\'{\i}}az}}]{2015MNRAS.451..888C}
{Casado}, J., {Ascasibar}, Y., {Gavil{\'a}n}, M., {et~al.} 2015, \mnras, 451,
  888

\bibitem[{{Chiaberge} {et~al.}(2015){Chiaberge}, {Gilli}, {Lotz}, \&
  {Norman}}]{2015ApJ...806..147C}
{Chiaberge}, M., {Gilli}, R., {Lotz}, J.~M., \& {Norman}, C. 2015, \apj, 806,
  147

\bibitem[{{Choi} {et~al.}(2009){Choi}, {Woo}, \& {Park}}]{2009ApJ...699.1679C}
{Choi}, Y.-Y., {Woo}, J.-H., \& {Park}, C. 2009, \apj, 699, 1679

\bibitem[{{Condon} {et~al.}(1998){Condon}, {Cotton}, {Greisen}, {Yin},
  {Perley}, {Taylor}, \& {Broderick}}]{1998AJ....115.1693C}
{Condon}, J.~J., {Cotton}, W.~D., {Greisen}, E.~W., {et~al.} 1998, \aj, 115,
  1693

\bibitem[{{Cowley} {et~al.}(2016){Cowley}, {Spitler}, {Tran}, {Rees},
  {Labb{\'e}}, {Allen}, {Brammer}, {Glazebrook}, {Hopkins}, {Juneau},
  {Kacprzak}, {Mullaney}, {Nanayakkara}, {Papovich}, {Quadri}, {Straatman},
  {Tomczak}, \& {van Dokkum}}]{2016MNRAS.457..629C}
{Cowley}, M.~J., {Spitler}, L.~R., {Tran}, K.-V.~H., {et~al.} 2016, \mnras,
  457, 629

\bibitem[{{Coziol} {et~al.}(2011){Coziol}, {Torres-Papaqui}, {Plauchu-Frayn},
  {Islas-Islas}, {Ortega-Minakata}, {Neri-Larios}, \&
  {Andernach}}]{2011RMxAA..47..361C}
{Coziol}, R., {Torres-Papaqui}, J.~P., {Plauchu-Frayn}, I., {et~al.} 2011,
  \rmxaa, 47, 361

\bibitem[{{Croton} {et~al.}(2006){Croton}, {Springel}, {White}, {De Lucia},
  {Frenk}, {Gao}, {Jenkins}, {Kauffmann}, {Navarro}, \&
  {Yoshida}}]{2006MNRAS.365...11C}
{Croton}, D.~J., {Springel}, V., {White}, S.~D.~M., {et~al.} 2006, \mnras, 365,
  11

\bibitem[{{Dressler}(1980)}]{1980ApJ...236..351D}
{Dressler}, A. 1980, \apj, 236, 351

\bibitem[{{Ellison} {et~al.}(2015){Ellison}, {Patton}, \&
  {Hickox}}]{2015MNRAS.451L..35E}
{Ellison}, S.~L., {Patton}, D.~R., \& {Hickox}, R.~C. 2015, \mnras, 451, L35

\bibitem[{{Ellison} {et~al.}(2011){Ellison}, {Patton}, {Mendel}, \&
  {Scudder}}]{2011MNRAS.418.2043E}
{Ellison}, S.~L., {Patton}, D.~R., {Mendel}, J.~T., \& {Scudder}, J.~M. 2011,
  \mnras, 418, 2043

\bibitem[{{Fabian}(1999)}]{1999MNRAS.308L..39F}
{Fabian}, A.~C. 1999, \mnras, 308, L39

\bibitem[{{Gr{\"u}tzbauch} {et~al.}(2011){Gr{\"u}tzbauch}, {Conselice},
  {Varela}, {Bundy}, {Cooper}, {Skibba}, \& {Willmer}}]{2011MNRAS.411..929G}
{Gr{\"u}tzbauch}, R., {Conselice}, C.~J., {Varela}, J., {et~al.} 2011, \mnras,
  411, 929

\bibitem[{{Hardcastle} {et~al.}(2007){Hardcastle}, {Evans}, \&
  {Croston}}]{2007MNRAS.376.1849H}
{Hardcastle}, M.~J., {Evans}, D.~A., \& {Croston}, J.~H. 2007, \mnras, 376,
  1849

\bibitem[{{Heckman}(1980)}]{1980A&A....87..152H}
{Heckman}, T.~M. 1980, \aap, 87, 152

\bibitem[{{Hern{\'a}ndez-Ibarra} {et~al.}(2013){Hern{\'a}ndez-Ibarra},
  {Dultzin}, {Krongold}, {Olmo}, {Perea}, \&
  {Gonz{\'a}lez}}]{2013MNRAS.434..336H}
{Hern{\'a}ndez-Ibarra}, F.~J., {Dultzin}, D., {Krongold}, Y., {et~al.} 2013,
  \mnras, 434, 336

\bibitem[{{Hern{\'a}ndez-Ibarra} {et~al.}(2016){Hern{\'a}ndez-Ibarra},
  {Krongold}, {Dultzin}, {del Olmo}, {Perea}, {Gonz{\'a}lez},
  {Mendoza-Castrej{\'o}n}, \& {Bitsakis}}]{2016MNRAS.tmp..262H}
{Hern{\'a}ndez-Ibarra}, F.~J., {Krongold}, Y., {Dultzin}, D., {et~al.} 2016,
  \mnras [\eprint[arXiv]{1509.02186}]

\bibitem[{{Hine} \& {Longair}(1979)}]{1979MNRAS.188..111H}
{Hine}, R.~G. \& {Longair}, M.~S. 1979, \mnras, 188, 111

\bibitem[{{Hong} {et~al.}(2015){Hong}, {Im}, {Kim}, \&
  {Ho}}]{2015ApJ...804...34H}
{Hong}, J., {Im}, M., {Kim}, M., \& {Ho}, L.~C. 2015, \apj, 804, 34

\bibitem[{Hunter(2007)}]{Hunter:2007}
Hunter, J.~D. 2007, Computing In Science \& Engineering, 9, 90

\bibitem[{Jones {et~al.}(2001)Jones, Oliphant, Peterson, {et~al.}}]{citescipy}
Jones, E., Oliphant, T., Peterson, P., {et~al.} 2001, {SciPy}: Open source
  scientific tools for {Python}, [Online; accessed 2016-01-15]

\bibitem[{{Karouzos} {et~al.}(2014){Karouzos}, {Im}, {Kim}, {Lee}, {Chapman},
  {Jeon}, {Choi}, {Hong}, {Hyun}, {Jun}, {Kim}, {Kim}, {Kim}, {Kim}, {Pak},
  {Park}, {Taak}, {Yoon}, \& {Edge}}]{2014ApJ...797...26K}
{Karouzos}, M., {Im}, M., {Kim}, J.-W., {et~al.} 2014, \apj, 797, 26

\bibitem[{{Kauffmann} {et~al.}(2003{\natexlab{a}}){Kauffmann}, {Heckman},
  {Tremonti}, {Brinchmann}, {Charlot}, {White}, {Ridgway}, {Brinkmann},
  {Fukugita}, {Hall}, {Ivezi{\'c}}, {Richards}, \&
  {Schneider}}]{2003MNRAS.346.1055K}
{Kauffmann}, G., {Heckman}, T.~M., {Tremonti}, C., {et~al.} 2003{\natexlab{a}},
  \mnras, 346, 1055

\bibitem[{{Kauffmann} {et~al.}(2003{\natexlab{b}}){Kauffmann}, {Heckman},
  {White}, {Charlot}, {Tremonti}, {Brinchmann}, {Bruzual}, {Peng}, {Seibert},
  {Bernardi}, {Blanton}, {Brinkmann}, {Castander}, {Cs{\'a}bai}, {Fukugita},
  {Ivezic}, {Munn}, {Nichol}, {Padmanabhan}, {Thakar}, {Weinberg}, \&
  {York}}]{2003MNRAS.341...33K}
{Kauffmann}, G., {Heckman}, T.~M., {White}, S.~D.~M., {et~al.}
  2003{\natexlab{b}}, \mnras, 341, 33

\bibitem[{{Kauffmann} {et~al.}(2004){Kauffmann}, {White}, {Heckman},
  {M{\'e}nard}, {Brinchmann}, {Charlot}, {Tremonti}, \&
  {Brinkmann}}]{2004MNRAS.353..713K}
{Kauffmann}, G., {White}, S.~D.~M., {Heckman}, T.~M., {et~al.} 2004, \mnras,
  353, 713

\bibitem[{{Kaviraj} {et~al.}(2015){Kaviraj}, {Shabala}, {Deller}, \&
  {Middelberg}}]{2015MNRAS.452..774K}
{Kaviraj}, S., {Shabala}, S.~S., {Deller}, A.~T., \& {Middelberg}, E. 2015,
  \mnras, 452, 774

\bibitem[{{Kere{\v s}} {et~al.}(2009){Kere{\v s}}, {Katz}, {Fardal},
  {Dav{\'e}}, \& {Weinberg}}]{2009MNRAS.395..160K}
{Kere{\v s}}, D., {Katz}, N., {Fardal}, M., {Dav{\'e}}, R., \& {Weinberg},
  D.~H. 2009, \mnras, 395, 160

\bibitem[{{Li} {et~al.}(2008){Li}, {Kauffmann}, {Heckman}, {White}, \&
  {Jing}}]{2008MNRAS.385.1915L}
{Li}, C., {Kauffmann}, G., {Heckman}, T.~M., {White}, S.~D.~M., \& {Jing},
  Y.~P. 2008, \mnras, 385, 1915

\bibitem[{{Liu} {et~al.}(2015){Liu}, {Pan}, {Hao}, {Hoyle}, {Constantin}, \&
  {Vogeley}}]{2015ApJ...810..165L}
{Liu}, C.-X., {Pan}, D.~C., {Hao}, L., {et~al.} 2015, \apj, 810, 165

\bibitem[{{Lynden-Bell}(1969)}]{1969Natur.223..690L}
{Lynden-Bell}, D. 1969, \nat, 223, 690

\bibitem[{{Magliocchetti} {et~al.}(2016){Magliocchetti}, {Lutz}, {Santini},
  {Salvato}, {Popesso}, {Berta}, \& {Pozzi}}]{2016MNRAS.456..431M}
{Magliocchetti}, M., {Lutz}, D., {Santini}, P., {et~al.} 2016, \mnras, 456, 431

\bibitem[{{Manzer} \& {De Robertis}(2014)}]{2014ApJ...788..140M}
{Manzer}, L.~H. \& {De Robertis}, M.~M. 2014, \apj, 788, 140

\bibitem[{{McAlpine} {et~al.}(2015){McAlpine}, {Prandoni}, {Jarvis}, {Seymour},
  {Padovani}, {Best}, {Simpson}, {Guidetti}, {Murphy}, {Huynh}, {Vaccari},
  {White}, {Beswick}, {Afonso}, {Magliocchetti}, \&
  {Bondi}}]{2015aska.confE..83M}
{McAlpine}, K., {Prandoni}, I., {Jarvis}, M., {et~al.} 2015, Advancing
  Astrophysics with the Square Kilometre Array (AASKA14), 83

\bibitem[{{Melnyk} {et~al.}(2015){Melnyk}, {Karachentseva}, \&
  {Karachentsev}}]{2015MNRAS.451.1482M}
{Melnyk}, O., {Karachentseva}, V., \& {Karachentsev}, I. 2015, \mnras, 451,
  1482

\bibitem[{{Pace} \& {Salim}(2014)}]{2014ApJ...785...66P}
{Pace}, C. \& {Salim}, S. 2014, \apj, 785, 66

\bibitem[{{Pan} {et~al.}(2012){Pan}, {Vogeley}, {Hoyle}, {Choi}, \&
  {Park}}]{2012MNRAS.421..926P}
{Pan}, D.~C., {Vogeley}, M.~S., {Hoyle}, F., {Choi}, Y.-Y., \& {Park}, C. 2012,
  \mnras, 421, 926

\bibitem[{{Peng} {et~al.}(2015){Peng}, {Maiolino}, \&
  {Cochrane}}]{2015Natur.521..192P}
{Peng}, Y., {Maiolino}, R., \& {Cochrane}, R. 2015, \nat, 521, 192

\bibitem[{{Peng} {et~al.}(2010){Peng}, {Lilly}, {Kova{\v c}}, {Bolzonella},
  {Pozzetti}, {Renzini}, {Zamorani}, {Ilbert}, {Knobel}, {Iovino}, {Maier},
  {Cucciati}, {Tasca}, {Carollo}, {Silverman}, {Kampczyk}, {de Ravel},
  {Sanders}, {Scoville}, {Contini}, {Mainieri}, {Scodeggio}, {Kneib}, {Le
  F{\`e}vre}, {Bardelli}, {Bongiorno}, {Caputi}, {Coppa}, {de la Torre},
  {Franzetti}, {Garilli}, {Lamareille}, {Le Borgne}, {Le Brun}, {Mignoli},
  {Perez Montero}, {Pello}, {Ricciardelli}, {Tanaka}, {Tresse}, {Vergani},
  {Welikala}, {Zucca}, {Oesch}, {Abbas}, {Barnes}, {Bordoloi}, {Bottini},
  {Cappi}, {Cassata}, {Cimatti}, {Fumana}, {Hasinger}, {Koekemoer},
  {Leauthaud}, {Maccagni}, {Marinoni}, {McCracken}, {Memeo}, {Meneux}, {Nair},
  {Porciani}, {Presotto}, \& {Scaramella}}]{2010ApJ...721..193P}
{Peng}, Y.-j., {Lilly}, S.~J., {Kova{\v c}}, K., {et~al.} 2010, \apj, 721, 193

\bibitem[{{Peng} {et~al.}(2012){Peng}, {Lilly}, {Renzini}, \&
  {Carollo}}]{2012ApJ...757....4P}
{Peng}, Y.-j., {Lilly}, S.~J., {Renzini}, A., \& {Carollo}, M. 2012, \apj, 757,
  4

\bibitem[{P\'erez \& Granger(2007)}]{PER-GRA:2007}
P\'erez, F. \& Granger, B.~E. 2007, Computing in Science and Engineering, 9, 21

\bibitem[{{Pulatova} {et~al.}(2015){Pulatova}, {Vavilova}, {Sawangwit},
  {Babyk}, \& {Klimanov}}]{2015MNRAS.447.2209P}
{Pulatova}, N.~G., {Vavilova}, I.~B., {Sawangwit}, U., {Babyk}, I., \&
  {Klimanov}, S. 2015, \mnras, 447, 2209

\bibitem[{{Ramos Almeida} {et~al.}(2012){Ramos Almeida}, {Bessiere},
  {Tadhunter}, {P{\'e}rez-Gonz{\'a}lez}, {Barro}, {Inskip}, {Morganti}, {Holt},
  \& {Dicken}}]{2012MNRAS.419..687R}
{Ramos Almeida}, C., {Bessiere}, P.~S., {Tadhunter}, C.~N., {et~al.} 2012,
  \mnras, 419, 687

\bibitem[{{Rosario} {et~al.}(2016){Rosario}, {Mendel}, {Ellison}, {Lutz}, \&
  {Trump}}]{2016MNRAS.457.2703R}
{Rosario}, D.~J., {Mendel}, J.~T., {Ellison}, S.~L., {Lutz}, D., \& {Trump},
  J.~R. 2016, \mnras, 457, 2703

\bibitem[{{Sabater} {et~al.}(2013){Sabater}, {Best}, \&
  {Argudo-Fern{\'a}ndez}}]{2013MNRAS.430..638S}
{Sabater}, J., {Best}, P.~N., \& {Argudo-Fern{\'a}ndez}, M. 2013, \mnras, 430,
  638

\bibitem[{{Sabater} {et~al.}(2015){Sabater}, {Best}, \&
  {Heckman}}]{2015MNRAS.447..110S}
{Sabater}, J., {Best}, P.~N., \& {Heckman}, T.~M. 2015, \mnras, 447, 110

\bibitem[{{Sabater} {et~al.}(2008){Sabater}, {Leon}, {Verdes-Montenegro},
  {Lisenfeld}, {Sulentic}, \& {Verley}}]{2008A&A...486...73S}
{Sabater}, J., {Leon}, S., {Verdes-Montenegro}, L., {et~al.} 2008, \aap, 486,
  73

\bibitem[{{Sabater} {et~al.}(2012){Sabater}, {Verdes-Montenegro}, {Leon},
  {Best}, \& {Sulentic}}]{2012A&A...545A..15S}
{Sabater}, J., {Verdes-Montenegro}, L., {Leon}, S., {Best}, P., \& {Sulentic},
  J. 2012, \aap, 545, A15

\bibitem[{{Salim} {et~al.}(2007){Salim}, {Rich}, {Charlot}, {Brinchmann},
  {Johnson}, {Schiminovich}, {Seibert}, {Mallery}, {Heckman}, {Forster},
  {Friedman}, {Martin}, {Morrissey}, {Neff}, {Small}, {Wyder}, {Bianchi},
  {Donas}, {Lee}, {Madore}, {Milliard}, {Szalay}, {Welsh}, \&
  {Yi}}]{2007ApJS..173..267S}
{Salim}, S., {Rich}, R.~M., {Charlot}, S., {et~al.} 2007, \apjs, 173, 267

\bibitem[{{Salpeter}(1964)}]{1964ApJ...140..796S}
{Salpeter}, E.~E. 1964, \apj, 140, 796

\bibitem[{{Satyapal} {et~al.}(2014){Satyapal}, {Ellison}, {McAlpine}, {Hickox},
  {Patton}, \& {Mendel}}]{2014MNRAS.441.1297S}
{Satyapal}, S., {Ellison}, S.~L., {McAlpine}, W., {et~al.} 2014, \mnras, 441,
  1297

\bibitem[{{Shakura} \& {Sunyaev}(1973)}]{1973A&A....24..337S}
{Shakura}, N.~I. \& {Sunyaev}, R.~A. 1973, \aap, 24, 337

\bibitem[{{Shen} {et~al.}(2016){Shen}, {Argudo-Fern{\'a}ndez}, {Chen}, {Chen},
  {Feng}, {Hou}, {Hou}, {Jiang}, {Jing}, {Kong}, {Luo}, {Luo}, {Shao}, {Wang},
  {Wang}, {Wang}, {Wu}, {Wu}, {Yang}, {Yang}, {Yuan}, {Yuan}, {Zhang}, {Zhang},
  \& {Zhang}}]{2016RAA....16c...7S}
{Shen}, S.-Y., {Argudo-Fern{\'a}ndez}, M., {Chen}, L., {et~al.} 2016, Research
  in Astronomy and Astrophysics, 16, 007

\bibitem[{{Soltan}(1982)}]{1982MNRAS.200..115S}
{Soltan}, A. 1982, \mnras, 200, 115

\bibitem[{{Strauss} {et~al.}(2002){Strauss}, {Weinberg}, {Lupton}, {Narayanan},
  {Annis}, {Bernardi}, {Blanton}, {Burles}, {Connolly}, {Dalcanton}, {Doi},
  {Eisenstein}, {Frieman}, {Fukugita}, {Gunn}, \v{Z}. {Ivezi{\'c}}, {Kent},
  {Kim}, {Knapp}, {Kron}, {Munn}, {Newberg}, {Nichol}, {Okamura}, {Quinn},
  {Richmond}, {Schlegel}, {Shimasaku}, {SubbaRao}, {Szalay}, {Vanden Berk},
  {Vogeley}, {Yanny}, {Yasuda}, {York}, \& {Zehavi}}]{2002AJ....124.1810S}
{Strauss}, M.~A., {Weinberg}, D.~H., {Lupton}, R.~H., {et~al.} 2002, \aj, 124,
  1810

\bibitem[{{Tasse} {et~al.}(2008){Tasse}, {Best}, {R{\"o}ttgering}, \& {Le
  Borgne}}]{2008A&A...490..893T}
{Tasse}, C., {Best}, P.~N., {R{\"o}ttgering}, H., \& {Le Borgne}, D. 2008,
  \aap, 490, 893

\bibitem[{{Taylor}(2005)}]{2005ASPC..347...29T}
{Taylor}, M.~B. 2005, in Astronomical Society of the Pacific Conference Series,
  Vol. 347, Astronomical Data Analysis Software and Systems XIV, ed.
  {P.~Shopbell, M.~Britton, \& R.~Ebert}, 29

\bibitem[{{Tremonti} {et~al.}(2004){Tremonti}, {Heckman}, {Kauffmann},
  {Brinchmann}, {Charlot}, {White}, {Seibert}, {Peng}, {Schlegel}, {Uomoto},
  {Fukugita}, \& {Brinkmann}}]{2004ApJ...613..898T}
{Tremonti}, C.~A., {Heckman}, T.~M., {Kauffmann}, G., {et~al.} 2004, \apj, 613,
  898

\bibitem[{{Verdes-Montenegro} {et~al.}(2005){Verdes-Montenegro}, {Sulentic},
  {Lisenfeld}, {Leon}, {Espada}, {Garcia}, {Sabater}, \&
  {Verley}}]{2005A&A...436..443V}
{Verdes-Montenegro}, L., {Sulentic}, J., {Lisenfeld}, U., {et~al.} 2005, \aap,
  436, 443

\bibitem[{{Verley} {et~al.}(2007){Verley}, {Leon}, {Verdes-Montenegro},
  {Combes}, {Sabater}, {Sulentic}, {Bergond}, {Espada}, {Garc{\'i}a},
  {Lisenfeld}, \& {Odewahn}}]{2007A&A...472..121V}
{Verley}, S., {Leon}, S., {Verdes-Montenegro}, L., {et~al.} 2007, \aap, 472,
  121

\bibitem[{{von der Linden} {et~al.}(2010){von der Linden}, {Wild}, {Kauffmann},
  {White}, \& {Weinmann}}]{2010MNRAS.404.1231V}
{von der Linden}, A., {Wild}, V., {Kauffmann}, G., {White}, S.~D.~M., \&
  {Weinmann}, S. 2010, \mnras, 404, 1231

\bibitem[{Walt {et~al.}(2011)Walt, Colbert, \&
  Varoquaux}]{:/content/aip/journal/cise/13/2/10.1109/MCSE.2011.37}
Walt, S. v.~d., Colbert, S.~C., \& Varoquaux, G. 2011, Computing in Science \&
  Engineering, 13, 22

\bibitem[{{Wild} {et~al.}(2010){Wild}, {Heckman}, \&
  {Charlot}}]{2010MNRAS.405..933W}
{Wild}, V., {Heckman}, T., \& {Charlot}, S. 2010, \mnras, 405, 933

\bibitem[{{Yang} {et~al.}(2007){Yang}, {Mo}, {van den Bosch}, {Pasquali}, {Li},
  \& {Barden}}]{2007ApJ...671..153Y}
{Yang}, X., {Mo}, H.~J., {van den Bosch}, F.~C., {et~al.} 2007, \apj, 671, 153

\bibitem[{{York} {et~al.}(2000){York}, {Adelman}, {Anderson}, {Anderson},
  {Annis}, {Bahcall}, {Bakken}, {Barkhouser}, {Bastian}, {Berman}, {Boroski},
  {Bracker}, {Briegel}, {Briggs}, {Brinkmann}, {Brunner}, {Burles}, {Carey},
  {Carr}, {Castander}, {Chen}, {Colestock}, {Connolly}, {Crocker}, {Csabai},
  {Czarapata}, {Davis}, {Doi}, {Dombeck}, {Eisenstein}, {Ellman}, {Elms},
  {Evans}, {Fan}, {Federwitz}, {Fiscelli}, {Friedman}, {Frieman}, {Fukugita},
  {Gillespie}, {Gunn}, {Gurbani}, {de Haas}, {Haldeman}, {Harris}, {Hayes},
  {Heckman}, {Hennessy}, {Hindsley}, {Holm}, {Holmgren}, {Huang}, {Hull},
  {Husby}, {Ichikawa}, {Ichikawa}, \v{Z}. {Ivezi{\'c}}, {Kent}, {Kim},
  {Kinney}, {Klaene}, {Kleinman}, {Kleinman}, {Knapp}, {Korienek}, {Kron},
  {Kunszt}, {Lamb}, {Lee}, {Leger}, {Limmongkol}, {Lindenmeyer}, {Long},
  {Loomis}, {Loveday}, {Lucinio}, {Lupton}, {MacKinnon}, {Mannery}, {Mantsch},
  {Margon}, {McGehee}, {McKay}, {Meiksin}, {Merelli}, {Monet}, {Munn},
  {Narayanan}, {Nash}, {Neilsen}, {Neswold}, {Newberg}, {Nichol}, {Nicinski},
  {Nonino}, {Okada}, {Okamura}, {Ostriker}, {Owen}, {Pauls}, {Peoples},
  {Peterson}, {Petravick}, {Pier}, {Pope}, {Pordes}, {Prosapio},
  {Rechenmacher}, {Quinn}, {Richards}, {Richmond}, {Rivetta}, {Rockosi},
  {Ruthmansdorfer}, {Sandford}, {Schlegel}, {Schneider}, {Sekiguchi}, {Sergey},
  {Shimasaku}, {Siegmund}, {Smee}, {Smith}, {Snedden}, {Stone}, {Stoughton},
  {Strauss}, {Stubbs}, {SubbaRao}, {Szalay}, {Szapudi}, {Szokoly}, {Thakar},
  {Tremonti}, {Tucker}, {Uomoto}, {Vanden Berk}, {Vogeley}, {Waddell}, {Wang},
  {Watanabe}, {Weinberg}, {Yanny}, \& {Yasuda}}]{2000AJ....120.1579Y}
{York}, D.~G., {Adelman}, J., {Anderson}, J. J.~E., {et~al.} 2000, \aj, 120,
  1579

\bibitem[{{Zhao} {et~al.}(2012){Zhao}, {Zhao}, {Chu}, {Jing}, \&
  {Deng}}]{2012RAA....12..723Z}
{Zhao}, G., {Zhao}, Y.-H., {Chu}, Y.-Q., {Jing}, Y.-P., \& {Deng}, L.-C. 2012,
  Research in Astronomy and Astrophysics, 12, 723

\end{thebibliography}

\begin{appendix}

\section{LSSGALPY} \label{Sec:App1}

\begin{landscape}
\begin{figure}
\centering
\includegraphics[width=1.1\textwidth]{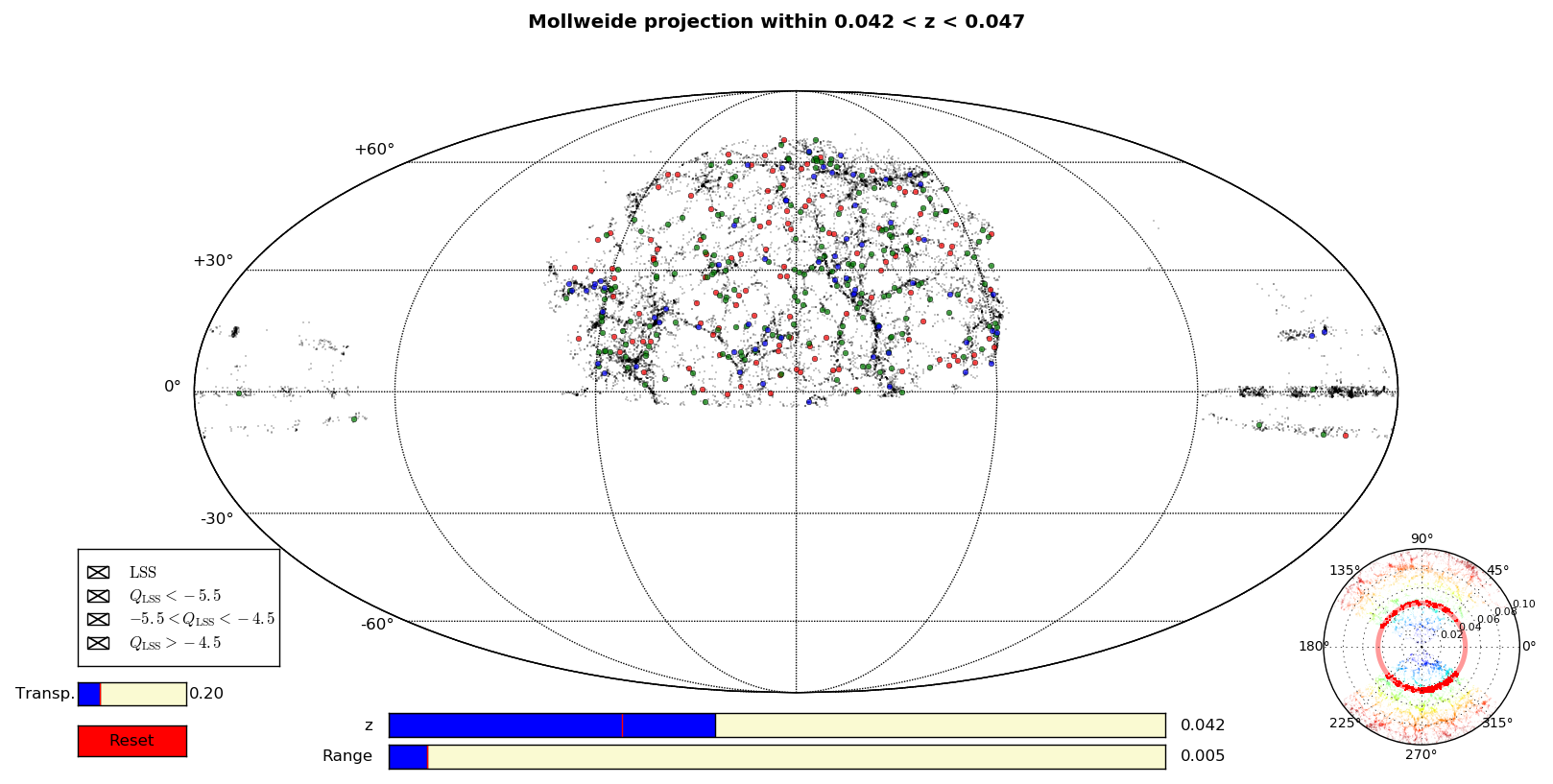}
\caption{Interactive 3D visualisation software: Mollweide projection. Mollweide projection of the LSS for galaxies (black points) in the redshift range $0.042 < z < 0.047$ as shown in the blue bars in the lower part of the figure. Red disks represent SIG galaxies with $Q_{\rm LSS}\leq-5.5$ within the same redshift range. Green disks represent SIG galaxies with $-5.5<Q_{\rm LSS}\leq-4.5$ within the same redshift range. Blue disks represent SIG galaxies with $Q_{\rm LSS}>-4.5$ within the same redshift range. To guide the eye, a wedge diagram, for LSS galaxies within -2 and 2 degrees in declination, is shown in the right lower part of the figure. Colour code according to the redshift from $z=0$ (blue) to $z=0.10$ (red). The red ring in the polar representation corresponds to the selected redshift range in the central Mollweide projection.}
\label{Fig:tool1}
\end{figure}
\end{landscape}

LSSGALPY\footnote{Available at \texttt{https://github.com/margudo/LSSGALPY}} \citep{2015ascl.soft05012A} is a tool for the interactive visualization of the large-scale environment around galaxies on the 3D space based on Python language. The tool allows one to easily compare the 3D positions of a sample (or samples) with respect to the locations of the LSS galaxies in their local and/or large scale environments. For the purpose of this study, we compared the position of galaxies in the SIG sample according to three different ranges of values of their $Q_{\rm LSS}$ (see a snapshot of the tool in Fig.~\ref{Fig:tool1}). 

We observe that most of the SIG galaxies with $Q_{\rm LSS}\leq-5.5$ are preferentially located in voids or low density regions, while SIG galaxies with $-5.5<Q_{\rm LSS}\leq-4.5$, and specially with $Q_{\rm LSS}>-4.5$, are distributed along the LSS. This means that these SIG galaxies mainly belong to the outer parts of filaments, walls, and clusters, and generally differ from the void population of galaxies.

\end{appendix}

\end{document}